\pdfoutput=1
\documentclass[journal]{IEEEtran}

\usepackage[noadjust]{cite}

\usepackage{color}

\usepackage{float}
\usepackage{graphicx}
\usepackage{epstopdf}
\usepackage{grffile}
\graphicspath{{./png/}{./jpeg/}}

\usepackage{multirow}

\usepackage{amsmath}
\usepackage{amssymb}
\usepackage{amsbsy}

\usepackage{algpseudocode}
\usepackage{algorithm}

\newcommand{\etal}{\textit{et al}}

\DeclareMathOperator*{\argmin}{argmin}

\begin{document}

\title{Object-based Multipass InSAR via Robust Low Rank Tensor Decomposition}

\author{Jian~Kang,~\IEEEmembership{Student~Member,~IEEE,}
	Yuanyuan~Wang,~\IEEEmembership{Member,~IEEE,}
	Michael Schmitt,~\IEEEmembership{Senior~Member,~IEEE,}
	and~Xiao~Xiang~Zhu~\IEEEmembership{Senior~Member,~IEEE}
	\thanks{This work is jointly supported by the China Scholarship Council, the European Research Council (ERC) under the European Union’s Horizon 2020 research and innovation programme (grant agreement No [ERC-2016-StG-714087], Acronym: \textit{So2Sat}), and Helmholtz Association under the framework of the Young Investigators Group ``SiPEO'' (VH-NG-1018, www.sipeo.bgu.tum.de)(\textit{Corresponding author: Xiao Xiang Zhu}).}
	\thanks{J. Kang, Y. Wang, M. Schmitt are with Signal Processing in Earth Observation (SiPEO), Technical University of Munich (TUM), 80333 Munich, Germany (e-mail: jian.kang@tum.de).}
	\thanks{X. X. Zhu is with the Remote Sensing Technology Institute (IMF), German Aerospace Center (DLR), 82234 Wessling, Germany, and also with Signal Processing in Earth Observation (SiPEO), Technical University of Munich (TUM), 80333 Munich, Germany (e-mail: xiao.zhu@dlr.de).}}
\markboth{IEEE Transactions on Geoscience and Remote Sensing, in press}%
{Shell \MakeLowercase{\textit{et al.}}: Bare Demo of IEEEtran.cls for IEEE Transactions on Magnetics Journals}

\IEEEtitleabstractindextext{
\begin{abstract}
	\textit{This is the pre-acceptance version, to read the final version please go to IEEE Transactions on Geoscience and Remote Sensing on IEEE Xplore.} 
	
	The most unique advantage of multipass SAR interferometry (InSAR) is the retrieval of long term geophysical parameters, e.g. linear deformation rates, over large areas. Recently, an object-based multipass InSAR framework has been proposed in \cite{kang2017robust}, as an alternative to the typical single-pixel methods, e.g. Persistent Scatterer Interferometry (PSI), or pixel-cluster-based methods, e.g. SqueeSAR. This enables the exploitation of inherent properties of InSAR phase stacks on an object level. As a follow-on, this paper investigates the inherent low rank property of such phase tensors, and proposes a \underline{Ro}bust \underline{M}ultipass \underline{I}nSAR technique via \underline{O}bject-based low rank tensor decomposition (RoMIO). We demonstrate that the filtered InSAR phase stacks can improve the accuracy of geophysical parameters estimated via conventional multipass InSAR techniques, e.g. PSI, by a factor of ten to thirty in typical settings. The proposed method is particularly effective against outliers, such as pixels with unmodeled phases. These merits in turn can effectively reduce the number of images required for a reliable estimation. The promising performance of the proposed method is demonstrated using high-resolution TerraSAR-X image stacks.

\end{abstract}
}

\maketitle
\IEEEdisplaynontitleabstractindextext
\IEEEpeerreviewmaketitle

\section{Introduction}\label{sec:introduction}
\subsection{Multipass InSAR}
\IEEEPARstart{M}{ultipass} or multibaseline InSAR techniques, such as persistent scatterer interferometry (PSI)
\cite{ferretti2001permanent,adam2003development,fornaro2009deformation,sousa2011persistent,gernhardt2012deformation,kampes2006radar,wang2014efficient,costantini2014persistent,zhang2011modeling,de2009detection}, distributed scatterer interferometry \cite{ferretti2011new,goel2012advanced,Wang201289,jiang2015fast,samiei2016phase,wang2016robust} and differential SAR tomography (D-TomoSAR) \cite{Fornaro2003,lombardini2005,zhu2010very,zhunolinear2011,reale2011,fornaro2014tomographic}, are the most popular methods for the retrieval of geophysical parameters (namely elevation and deformation parameters) for extended areas.

Past research on multipass InSAR was mainly focused on the optimal retrieval of the phase history parameters of individual scatterers, which can be considered in two categories: single-pixel-based methods and pixel-cluster-based methods. On one hand, single-pixel-based methods, such as PSI \cite{ferretti2001permanent,adam2003development,fornaro2009deformation,sousa2011persistent,gernhardt2012deformation,kampes2006radar,wang2014efficient,costantini2014persistent}, and D-TomoSAR \cite{Fornaro2003,lombardini2005,zhu2010very,zhunolinear2011,reale2011,fornaro2014tomographic}, have been widely applied to the monitoring of urban areas. In particular, significant development has been made in D-TomoSAR, such as super-resolution D-TomoSAR methods based on Compressive Sensing (CS) \cite{zhu2010tomographic,budillon2011three,zhu2014superresolving}, and combining D-TomoSAR with SAR geodesy \cite{eineder2011imaging,gisinger2015precise} to obtain absolute \textit{Geodetic TomoSAR} \cite{zhu2016geodetic} point clouds. On the other hand, pixel-cluster-based methods, such as SqueeSAR \cite{ferretti2011new,Wang201289,goel2012advanced,samiei2016phase,cao2016phase,schmitt2014adaptive,schmitt2014adaptivemultilooking}, CAESAR \cite{fornaro2015caesar} and TomoSAR based on distributed scatterers \cite{neumann2010estimation,tebaldini2010single,schmitt2014maximum}, exploit statistical similarities between the neighboring pixels, in order to retrieve the phase history parameters from their associated covariance matrices. Statistical ergodicity of the selected pixel clusters is always assumed in these methods for the estimation of the required sample covariance matrix. Likewise, nonlocal-InSAR (NL-InSAR) \cite{deledalle2011nl,zhu2014improving,deledalle2014exploiting,sica2015nonlocal} also selects similar pixels but based on patch similarity.

Although some of the abovementioned techniques do exploit information from multiple neighbouring pixels or patches, no explicit semantic and geometric information that might be preserved in the images has been utilized. In \cite{zhumslimmer}, Zhu \etal. demonstrated that by introducing building footprints from OpenStreetMap (OSM) as prior knowledge of pixels sharing similar heights into frameworks based on joint sparse reconstruction techniques, a highly accurate tomographic reconstruction can be achieved using only six interferograms, instead of the typically-required 20-100. Inspired by this, we recently proposed a general framework for object-based InSAR deformation reconstruction based on a tensor-model with a regularization term, which is combined with semantic information shown in SAR images, i.e. classification labels of different objects like bridges, roofs and fa\c{c}ades, for an improvement of deformation retrieval  \cite{kang2016object,kang2017robust}.

Based on the previous work, this paper seeks to investigate the inherent low rank property of multipass InSAR phase tensors, given semantic prior knowledge of objects. We propose a novel robust tensor decomposition method using iterative reweighting to recover an outlier-free phase stack for the retrieval of the geophysical parameters.

\subsection{Low rank modeling}
Low rank modeling has been applied in many research fields of data analysis, since high-dimensional data are often embedded in a low-dimensional subspace \cite{zhou2015low}. One of the best known low rank modeling approaches is Principle Component Analysis (PCA) \cite{jolliffe2002principal}, which finds a low rank version of the matrix by minimizing the approximation error to the original data matrix in a least-squares sense. It has been utilized for tackling various problems in remote sensing, such as SAR-image-based change detection \cite{yousif2013improving}, hyperspectral image denoising \cite{chen2011denoising}, data feature extraction \cite{yao2003genetic}, and so on. For applications in the InSAR field, PCA has recently been utilized for decomposing the scatterer covariance matrix in CAESAR \cite{fornaro2015caesar}, in order to separate layovered scatterers within individual pixels.

However, due to the assumption of independently and identically distributed (i.i.d.) Gaussian samples, PCA is sensitive to the existence of outliers. To robustly recover the low rank data matrix, \cite{candes2011robust} proposed Robust PCA (RPCA) to decompose the original matrix into a low rank data matrix and a sparse outlier matrix. For instance, RPCA was deployed for hyperspectral image restoration in \cite{zhang2014hyperspectral}, and a RPCA-based approach for separating stationary and moving targets in SAR imaging was investigated in \cite{borcea2013synthetic}. To deal with the data in a multidimensional case, \cite{goldfarb2014robust} proposed a robust low-rank tensor recovery method called Higher order RPCA (HoRPCA), which has been employed in our previous work \cite{kang2017robust} as an outlier filtering step for object-based InSAR deformation reconstruction.

\subsection{Contributions of this paper}
To this end, the contributions of this paper are three-fold:
\begin{itemize}
	\item Based on the tensor model of object-based InSAR phase stacks \cite{kang2017robust}, we study their multidimensional low rank property.
	\item With this prior knowledge, we propose a novel InSAR phase tensor low rank decomposition method using iterative reweighting, which is named as RoMIO.
	\item Using simulation and real data, we demonstrate that the InSAR phase stacks filtered by RoMIO can improve the accuracy of geophysical parameters estimated via conventional multipass InSAR techniques, e.g. PSI, by a factor of ten to thirty in typical settings, especially in the existence of outliers.
\end{itemize}

\subsection{Structure of this paper}
The rest of this paper is organized as follows. Section \ref{sc:low_rank_study} studies the low rank property of such phase stacks. In Section \ref{sc:RIRT}, the proposed RoMIO method is demonstrated for robustly recovering object-based InSAR phases. Experiments including simulated and real InSAR data are conducted to substantiate the performance of the proposed algorithm in Section \ref{sc:experiments}. We discuss the experimental results in Section \ref{sc:discussions}. Section \ref{sc:conclustion} draws the conclusion of this paper.

\section{Low rank property of InSAR phase stacks}\label{sc:low_rank_study}
\subsection{Tensor basics}
\begin{table*}
	\caption{Mathematic notations}
	\centering
	\renewcommand{\arraystretch}{1.3}
	\begin{tabular}{l | l}
		\hline
		$ \mathcal{X}, \mathbf{X}, \mathbf{x}, x $				& tensor, matrix, vector, scalar \\
		$ \mathbf{X}_{(n)} $									& mode-$ n $ unfolding of tensor $ \mathcal{X} $ \\
		$ (R_1,R_2,\cdots,R_N) $								& tensor multilinear rank, where $ R_n=\mathrm{Rank}(\mathbf{X}_{(n)}),\quad n=1,2,\cdots,N $ \\
		$ \langle\mathcal{X},\mathcal{Y}\rangle $				& inner product of tensor $ \mathcal{X} $ and $ \mathcal{Y} $, i.e. the sum of product of their entries \\
		$ \|\mathcal{X}\|_F $									& Frobenius norm of tensor $ \mathcal{X} $, i.e. $ \|\mathcal{X}\|_F=\sqrt{\langle\mathcal{X},\mathcal{X}\rangle} $ \\
		$ \mathrm{vec}(\mathcal{X}) $							& vectorization of $\mathcal{X}$ \\
		$ \|\mathcal{X}\|_1 $									& $ L_1 $ norm of tensor $ \mathcal{X} $, i.e. $ \|\mathcal{X}\|_1=\|\mathrm{vec}(\mathcal{X})\|_1 $ \\
		$ \|\mathbf{X}\|_* $									& matrix nuclear norm: the sum of its singular values, i.e. \(\|\mathbf{X}\|_*:=\sum_{i}\sigma_{i}\) \\
		$ \mathcal{Y}=\mathcal{X}\times_n \mathbf{A} $			& mode-$ n $ multiplication of tensor $ \mathcal{X} $ and matrix $ \mathbf{A} $, i.e. $ \mathbf{Y}_{(n)}=\mathbf{A}\mathbf{X}_{(n)} $ \\
		$ \otimes $                                             & outer product \\
		$ \odot $                                               & element-wise product \\
		\hline

	\end{tabular}
	\label{table:Mathematic_notation}

\end{table*}

A tensor can be considered as a multi-dimensional array. The \textit{order} of a tensor is the number of its \textit{modes} or \textit{dimensions}. A tensor of order $ N $ in the complex domain can be denoted as $ \mathcal{X} \in \mathbb{C}^{I_1 \times I_2 \cdots \times I_N}$ and its entries as $ x_{i_1,i_2,\cdots,i_N} $. Specifically, vector $ \mathbf{x} $ is a tensor of order one, and matrix $ \mathbf{X} $ can be represented as a tensor of order two. \textit{Fibers} are the higher-order analogy of matrix rows and columns, which are defined by fixing every index but one. \textit{Slices} of a tensor are obtained by fixing all but two indices. Matricization, also known as \textit{unfolding}, is the process of reordering the elements of a tensor into a matrix. Specifically, the mode-$ n $ unfolding of tensor $ \mathcal{X} $ is defined by $ \mathbf{X}_{(n)} $ that is obtained by arranging the mode-$ n $ fibers as the columns of the matrix. The utilized tensor notations are summarized in Table \ref{table:Mathematic_notation}. The detailed introductions about multilinear algebra are presented in \cite{kolda2009tensor,cichocki2015tensor}.

\subsection{Tensor model of object-based multipass InSAR phase stacks}

\begin{figure}
	\centering
	\includegraphics[width=0.45\textwidth]{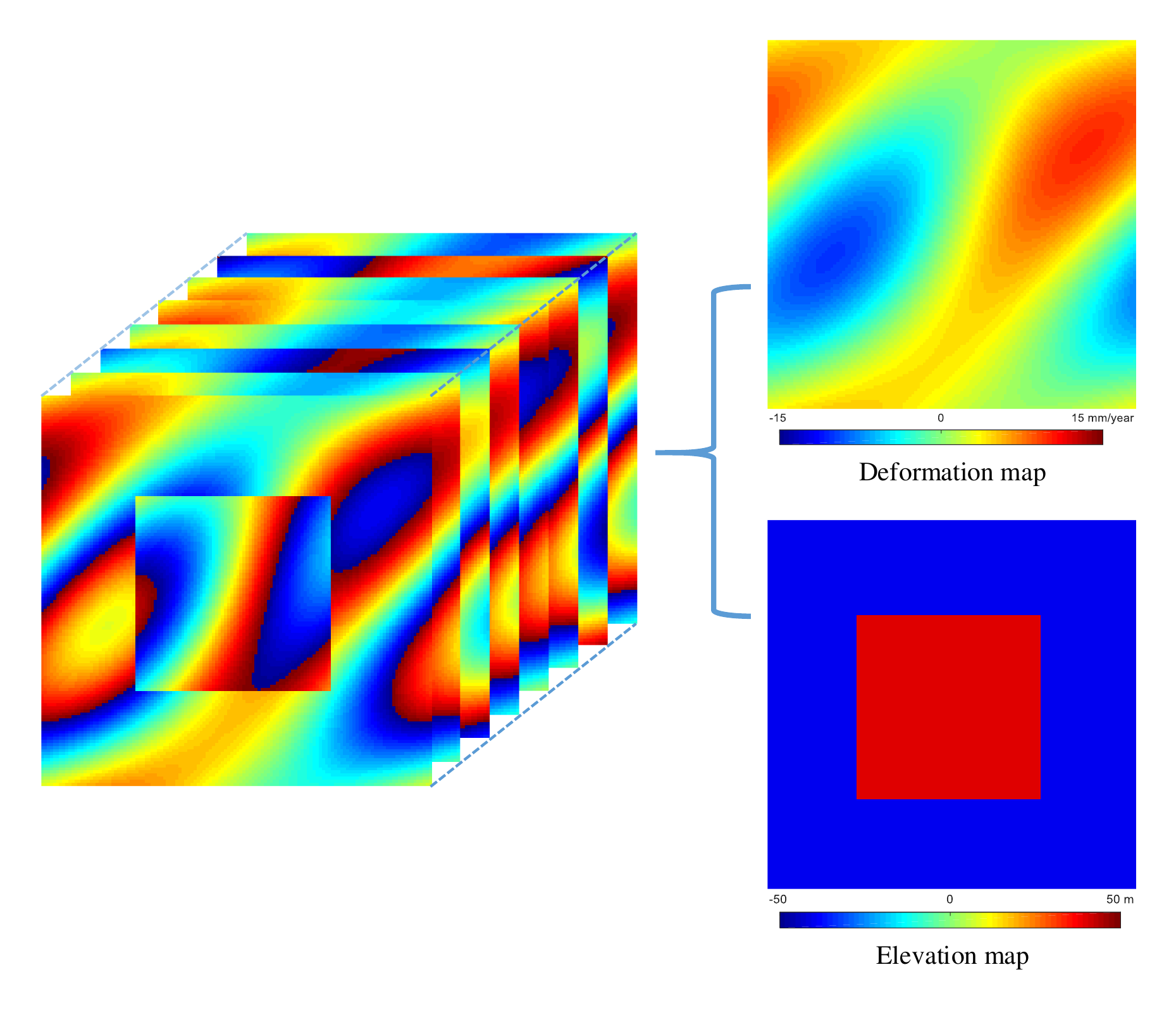}
	\caption{One example of an object-based InSAR phase stack, which can be represented by the tensor model in \eqref{eq:1}. It shows the wrapped phase stack, simulated by the synthetic linear deformation rates and elevations present on its right. The pattern of the simulated elevation map is comparable to that of urban objects in real scenarios. The simulated deformation map shows a more complex pattern, which represents continuously varying displacement in the scene. The elevation and deformation maps are designed to be spatially uncorrelated.}
	\label{fg:simu_InSAR_phase_stack}

\end{figure}

As proposed in our previous work \cite{kang2017robust}, given object areas, such as fa\c{c}ades, bridges or roofs, an InSAR phase stack can be represented by a 3-mode tensor: $ \mathcal{G}\in \mathbb{C}^{I_{1}\times{I_{2}}\times{I_{3}}} $, where $ I_{1} $ and $ I_{2} $ represent the spatial dimensions in range and azimuth, and $ I_3 $ denotes the number of SAR images. The phase tensor of the object can be modeled by
\begin{equation}
\overline{\mathcal{G}}(\mathbf{S},\mathbf{P})=\exp\{-j(\frac{4\pi}{\lambda{r}}\mathbf{S}\otimes\mathbf{b}+\frac{4\pi}{\lambda}\mathbf{P}\otimes\boldsymbol{\tau})\},
\label{eq:1}
\end{equation}
where $ \overline{\mathcal{G}} $ is the modeled phase tensor of the object, $ \mathbf{b}\in\mathbb{R}^{I_3} $ is the vector of the spatial baselines,  $ \boldsymbol{\tau}\in\mathbb{R}^{I_3} $ is a warped time variable \cite{zhunolinear2011}, e.g. $ \boldsymbol{\tau}=\mathbf{t} $ for a linear motion, and $ \boldsymbol{\tau}=\sin(2\pi(\mathbf{t}-t_0)) $ for a seasonal motion model with temporal baseline $ \mathbf{t} $ and time offset $ t_0 $. $ \mathbf{S}\in \mathbb{R}^{I_1\times I_2} $ and $ \mathbf{P}\in \mathbb{R}^{I_1\times I_2} $ are the unknown elevation and deformation maps to be estimated, respectively, $ \lambda $ is the wavelength of the radar signals and $ r $ denotes the range between radar and the observed object. The symbol $ \otimes $ denotes the outer product \cite{cichocki2015tensor}. A simulated example of such a phase stack is illustrated in Figure \ref{fg:simu_InSAR_phase_stack}. It shows the wrapped phase stack, and the simulated linear deformation rates and elevations from which the phase stack is constructed. The pattern of the simulated elevation map is comparable to that of urban objects in real scenarios. The simulated deformation map shows a more complex pattern, which represents continuously varying displacement in the scene. The elevation and deformation maps are designed to be spatially uncorrelated.

Such phase tensors in urban areas usually experience an inherent low rank nature, since it can be generally assumed that $ \mathbf{S} $ and $ \mathbf{P} $ follow certain regular structure or homogeneous pattern, because of the regular man-made structures in urban areas. Moreover, the observed SAR images of urban object areas are usually highly correlated along the temporal dimension. Such low rank property will be demonstrated and investigated in the following chapter.

\subsection{Low rank study of InSAR phase stacks}
Since PCA is the most basic low rank decomposition method for matrices, it will be employed in this section to demonstrate the low rank property of InSAR phase tensor. PCA is usually realized by Singular Value Decomposition (SVD) \cite{de1994singular}. Given a matrix $ \mathbf{X}\in\mathbb{C}^{I_1\times I_2} $ and its SVD, i.e. $ \mathbf{U}\mathbf{S}\mathbf{V}^H $, the rank $ R $ approximation of $ \mathbf{X} $ by truncating $ \mathbf{S} $ up to $ R $ dominant singular values is the matrix $ \mathbf{X}_R=\mathbf{U}_R\mathbf{S}_R\mathbf{V}_R^H $, where the $ R\times R $ diagonal matrix $ \mathbf{S}_R $ satisfies $ \mathbf{S}_R(i,i)=\mathbf{S}(i,i),i=1,2,\cdots,R $, $ \mathbf{U}_R $ is composed by the first $ R $ columns of $ \mathbf{U} $, and $ \mathbf{V}_R^H $ consists of the first $ R $ rows of $ \mathbf{V}^H $. This is also known as truncated SVD.

\begin{figure}
	\centering
	\includegraphics[scale=0.4]{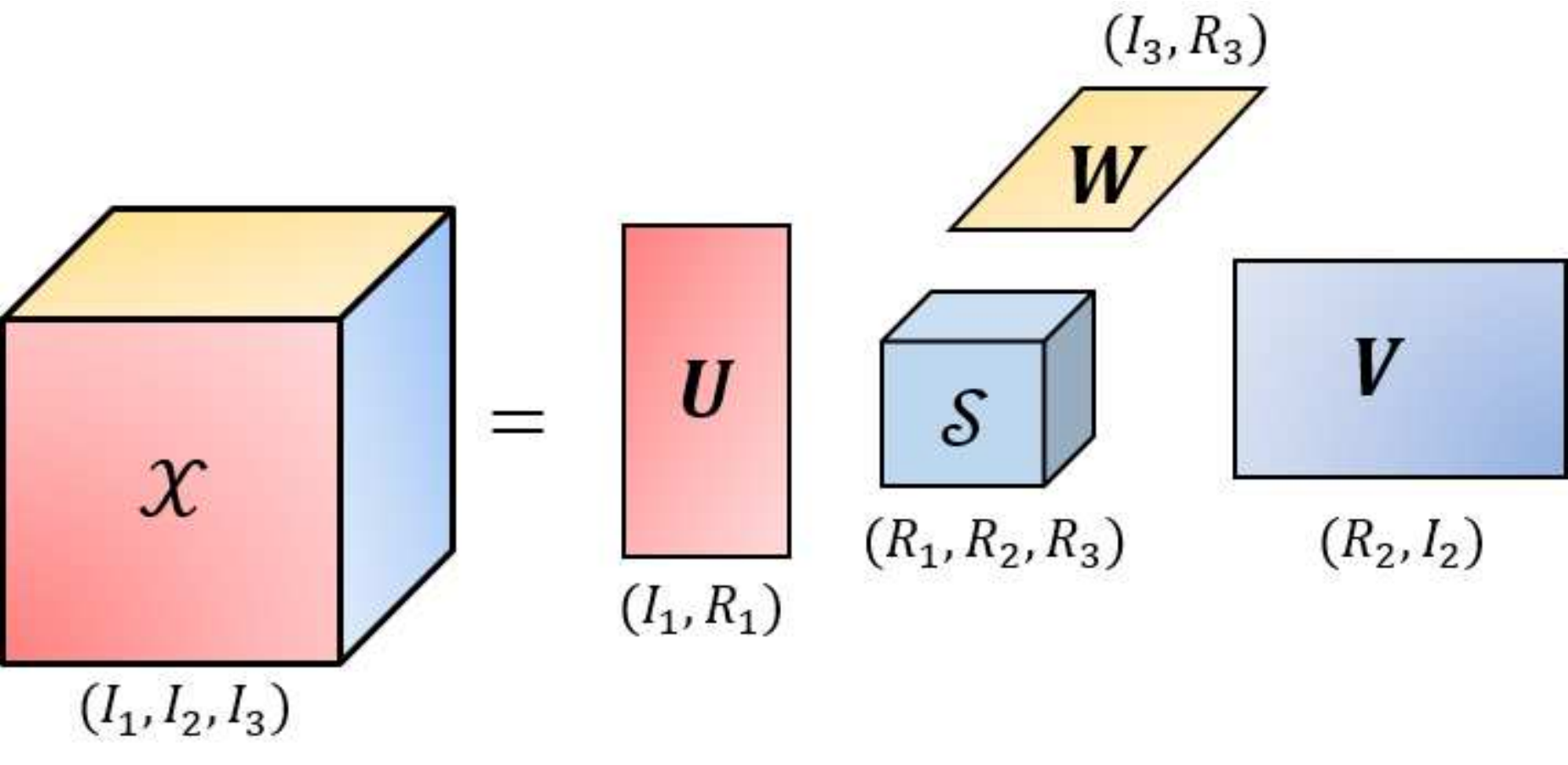}
	\caption{Illustration of Higher order Singular Value Decomposition (HoSVD) of a 3-mode tensor \cite{cichocki2015tensor}}
	\label{fg:HoSVD}
\end{figure}

As a higher-dimensional extension of SVD, Higher order Singular Value Decomposition (HoSVD), also known as Tucker decomposition \cite{de2000multilinear}, can provide a tensor data compression based on the low rank approximation, as illustrated in Figure \ref{fg:HoSVD}. It decomposes a tensor into a core tensor multiplied by a matrix along each mode. Specifically, for a 3-mode tensor, $ \mathcal{X}^{I_{1}\times{I_{2}}\times{I_{3}}} $, we have
\begin{equation}
\mathcal{X}=\mathcal{S}\times_1 \mathbf{U}\times_2\mathbf{V}\times_3\mathbf{W},
\label{eq:2}
\end{equation}
where $ \mathbf{U}^{I_1\times R_1} $, $ \mathbf{V}^{I_2\times R_2} $, and $ \mathbf{W}^{I_3\times R_3} $ are the factor matrices that can be considered as the principle components in each mode \cite{kolda2009tensor}, $ \mathcal{S}^{R_1\times R_2 \times R_3} $ is the so-called \textit{core tensor}, and symbol $ \times_n $ is mode-$ n $ multiplication between tensor and matrix \cite{cichocki2015tensor}. $ (R_1,R_2,R_3) $ is the so-called \textit{multilinear rank} of $ \mathcal{X} $. They fulfill the inequalities $ R_1\leqslant \min(I_1,I_2I_3),R_2\leqslant \min(I_2,I_1I_3), $ and $ R_3\leqslant \min(I_3,I_1I_2) $.

A low rank approximation of $ \mathcal{X} $ can be realized by the truncated HoSVD. Take $ \mathcal{X}^{I_{1}\times{I_{2}}\times{I_{3}}} $ as an example, we can define its tensor approximation with multilinear rank $ (K_1, K_2, K_3) $, where $ K_1\leqslant R_1,K_2\leqslant R_2,K_3\leqslant R_3 $, by the following truncated HoSVD:
\begin{equation}
\mathcal{X}^{I_{1}\times{I_{2}}\times{I_{3}}}\approx\mathcal{S}^{K_1\times K_2\times K_3}\times_1\mathbf{U}^{I_1\times K_1}\times_2\mathbf{V}^{I_2\times K_2}\times_3\mathbf{W}^{I_3\times K_3},
\label{eq:HoSVD_LR_approx}
\end{equation}
where $ \mathbf{U}^{I_1\times K_1}, \mathbf{V}^{I_2\times K_2} $ and $ \mathbf{W}^{I_3\times K_3} $ are created by storing the first $ K_i (i=1,2,3) $ singular vectors of $ \mathbf{U}, \mathbf{V} $ and $ \mathbf{W} $ and replacing the left $ R_i-K_i (i=1,2,3) $ vectors by zeros, and $ \mathcal{S}^{K_1\times K_2\times K_3} $ is created in a similar way. Such truncated HoSVD finds a low rank tensor approximation of the original tensor $ \mathcal{X} $ in a least-squares sense.

\begin{figure*}
	\centering
	\includegraphics[width=0.32\textwidth]{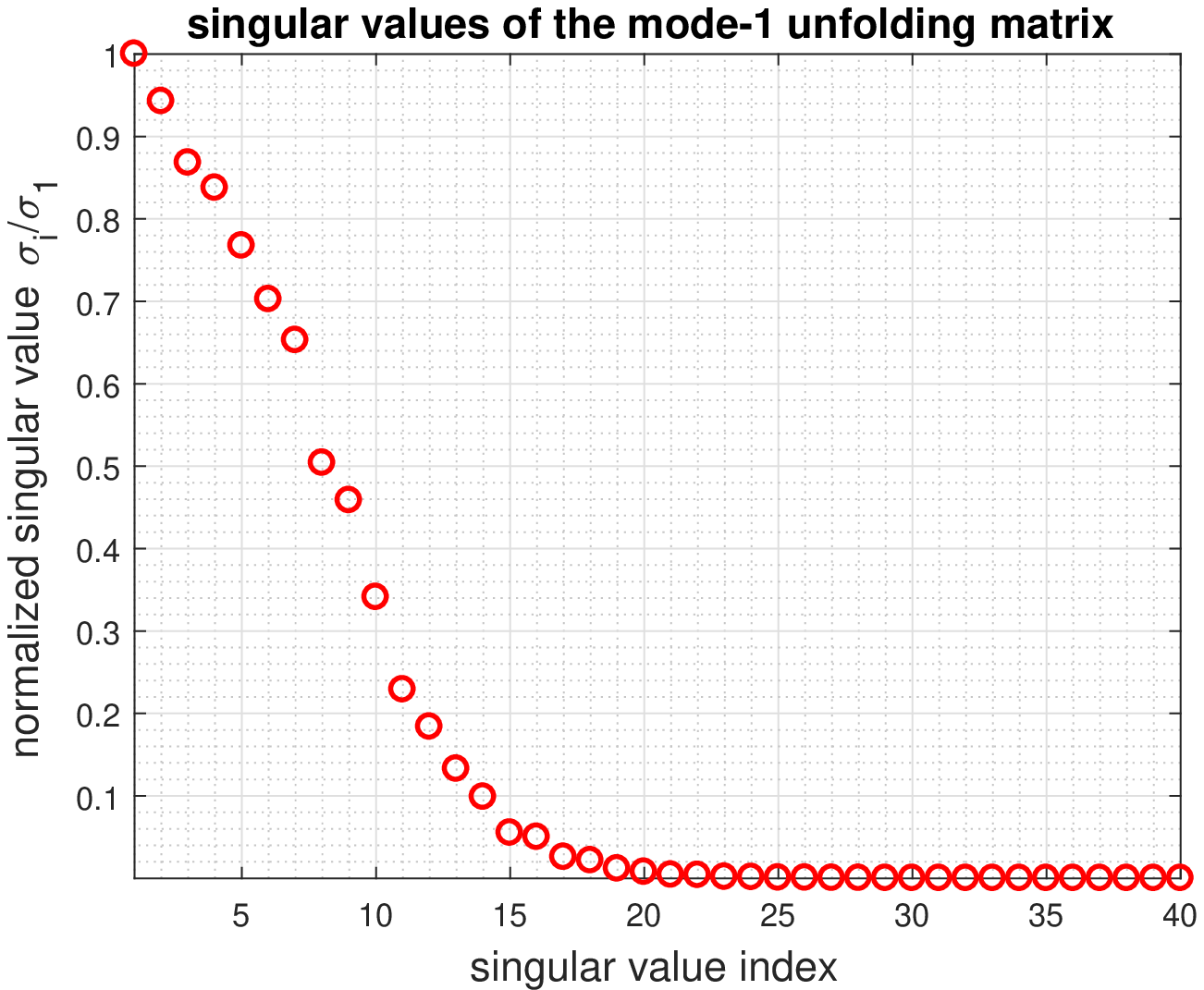}
	~
	\includegraphics[width=0.32\textwidth]{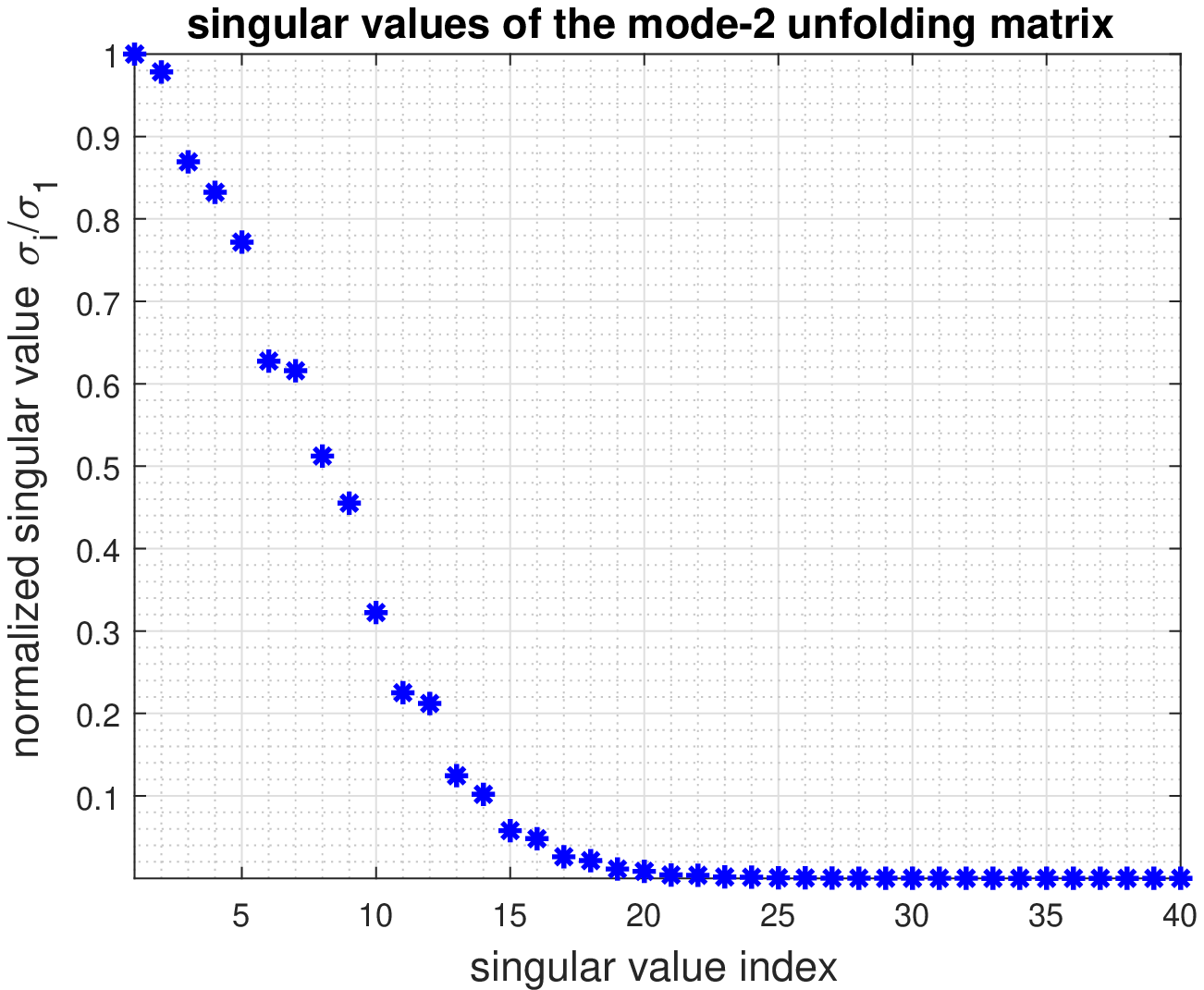}
	~
	\includegraphics[width=0.32\textwidth]{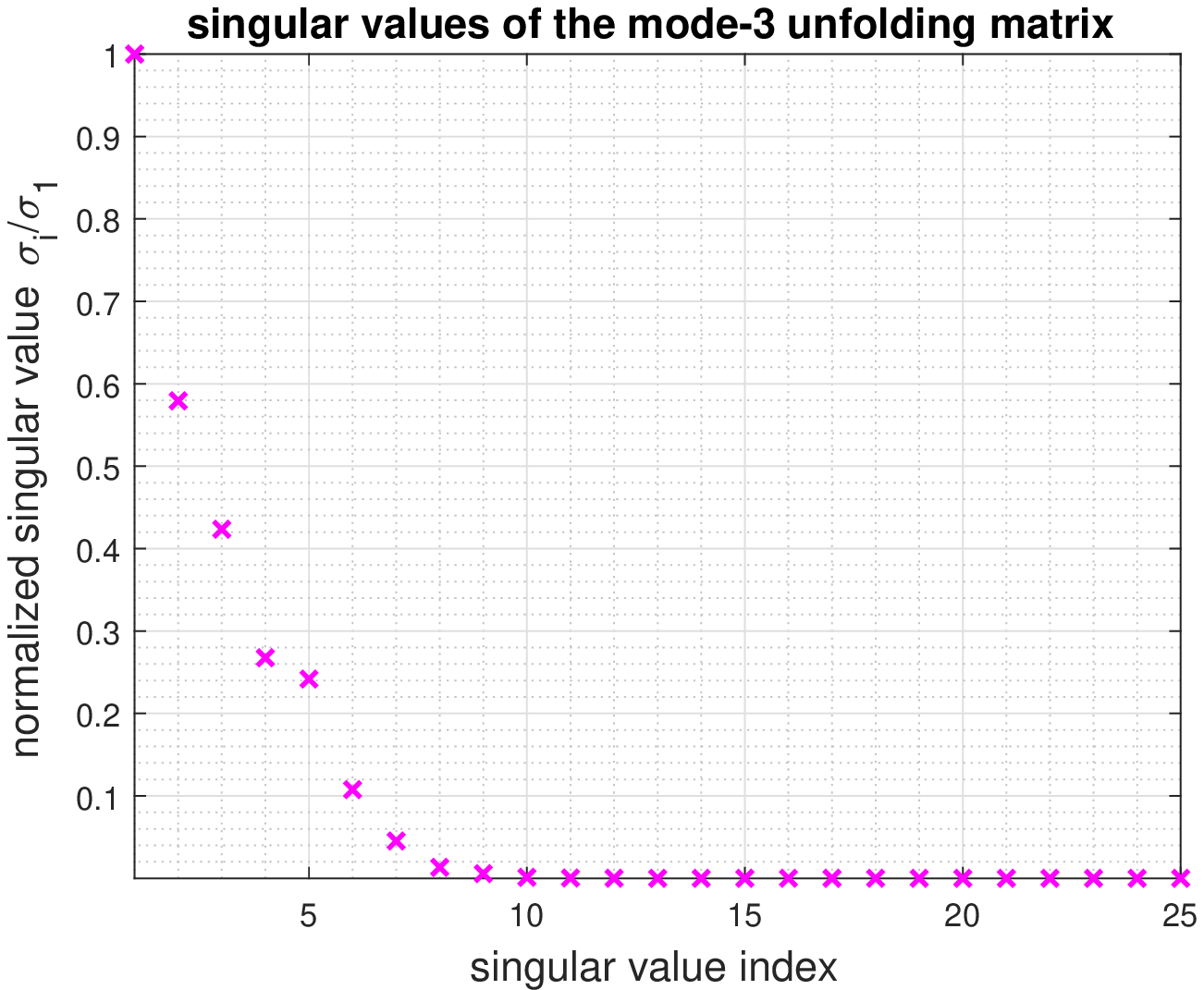}
	\caption{Plots of the normalized singular values of mode-1, -2 and -3 unfolding matrices of the simulated example of the complex-valued InSAR phase stack shown in Figure \ref{fg:simu_InSAR_phase_stack}. For visualization, we just plot the first $ 40 $ out of all the $ 128 $ normalized singular values of mode-1 and -2 unfolding matrices. It is demonstrated that the singular values of the three unfolding matrices decay rapidly, which indicates the low rank structure of the original tensor.}
	\label{fg:mode-1,2,3_SV}
\end{figure*}

\begin{figure}
	\centering
	\includegraphics[width=0.45\textwidth]{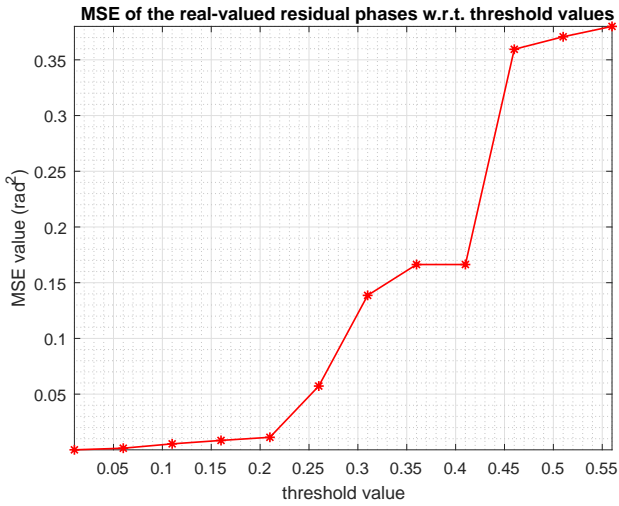}
	\caption{The Mean Square Error (MSE) values of the real-valued residual phases between the low rank approximated tensor $ \tilde{\mathcal{G}} $ and the original tensor $ \overline{\mathcal{G}} $ i.e. $ \mathrm{MSE}(\mathrm{angle}(\tilde{\mathcal{G}}\odot\mathrm{conj}(\overline{\mathcal{G}}))) $ w.r.t different threshold values.}
	\label{fg:MSE_compres_ratio_threshold}
\end{figure}

\begin{figure}
	\centering
	\includegraphics[width=0.45\textwidth]{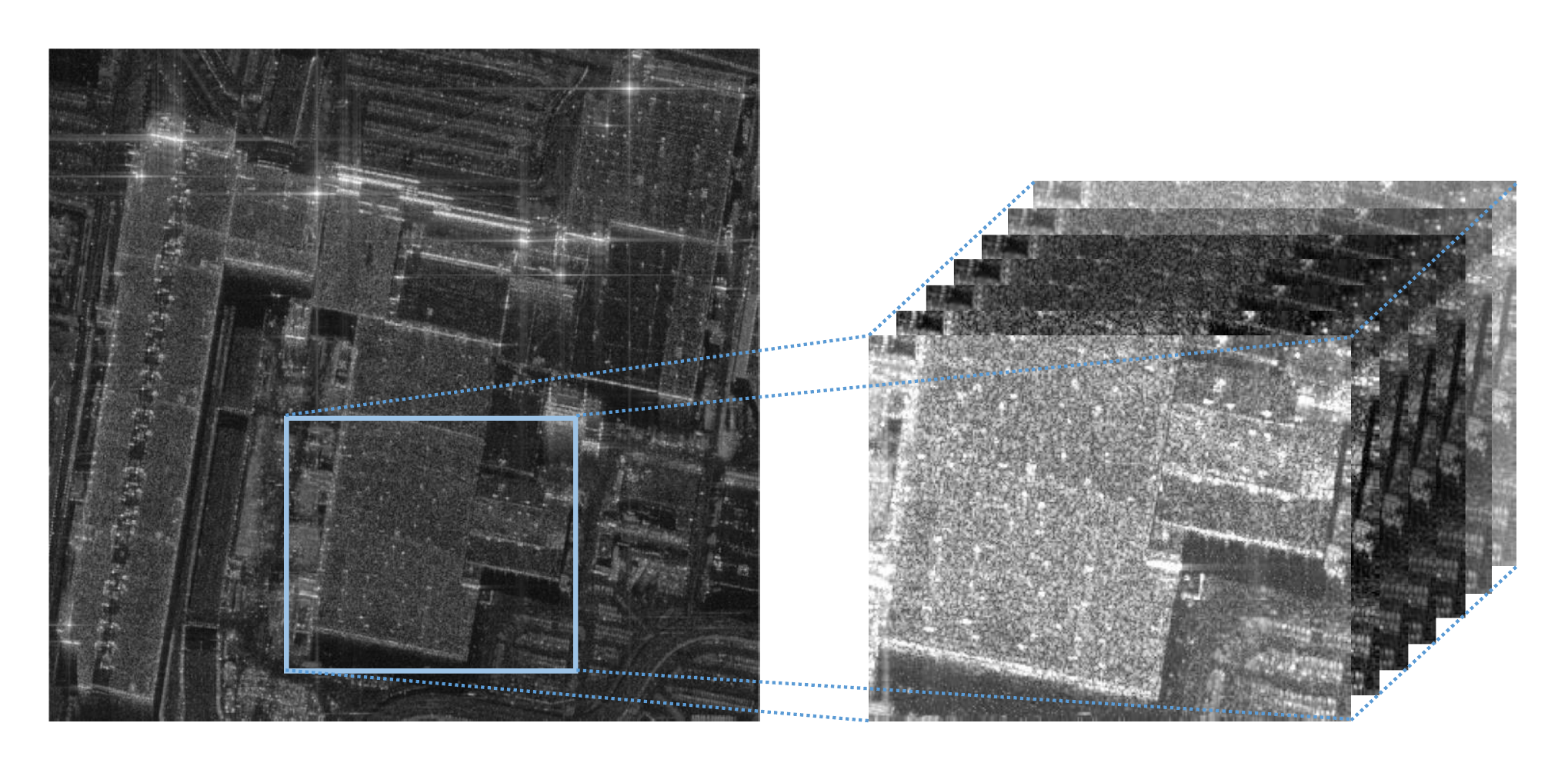}
	\caption{An InSAR phase tensor example of TerraSAR-X data with a roof area (blue rectangle) of LasVegas convention center. For the illustration, we show the amplitudes of the multipass SAR images.}
	\label{fg:roof_area_singular_values}

\end{figure}

\begin{figure*}
	\centering
	\includegraphics[width=0.32\textwidth]{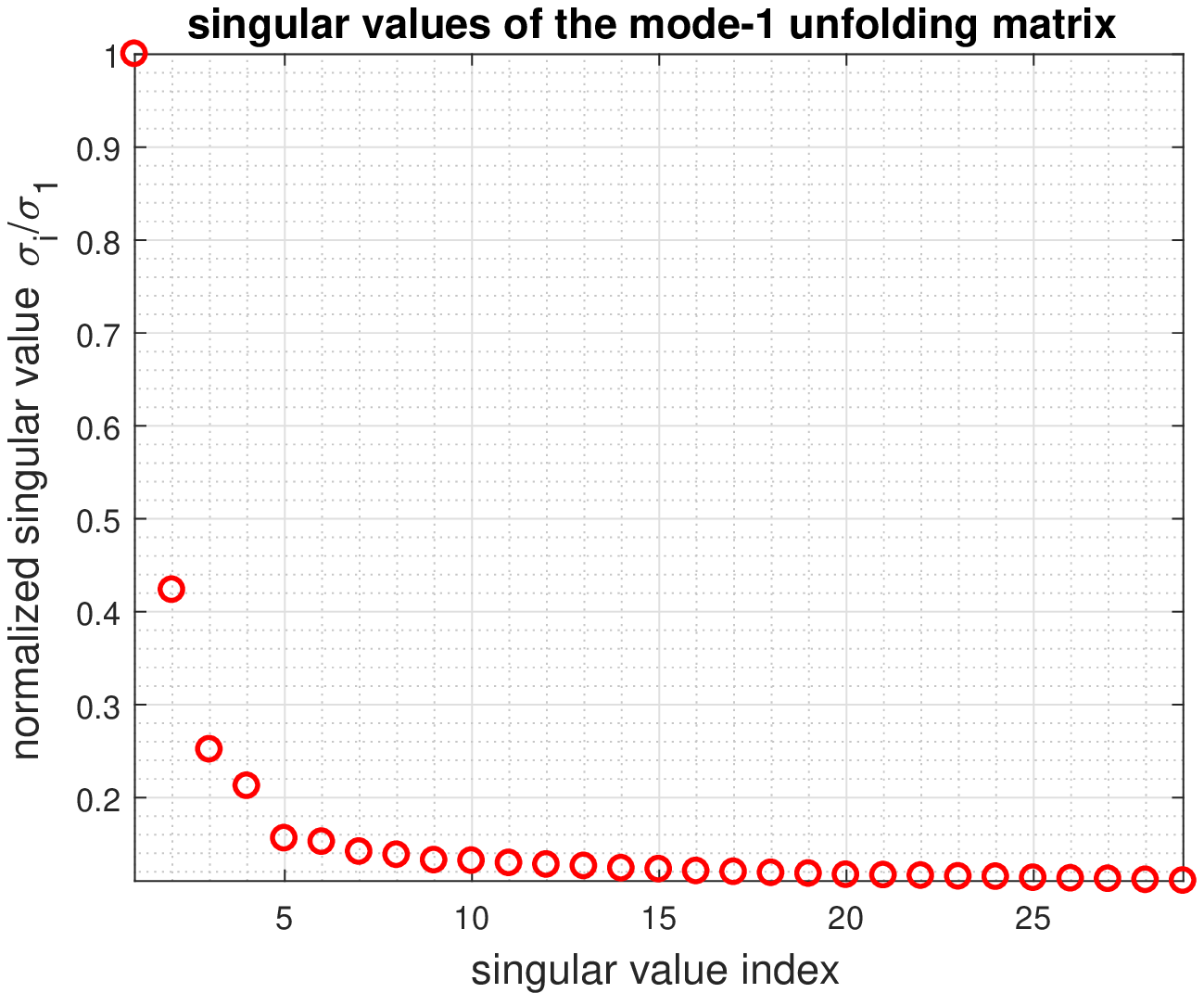}
	~
	\includegraphics[width=0.32\textwidth]{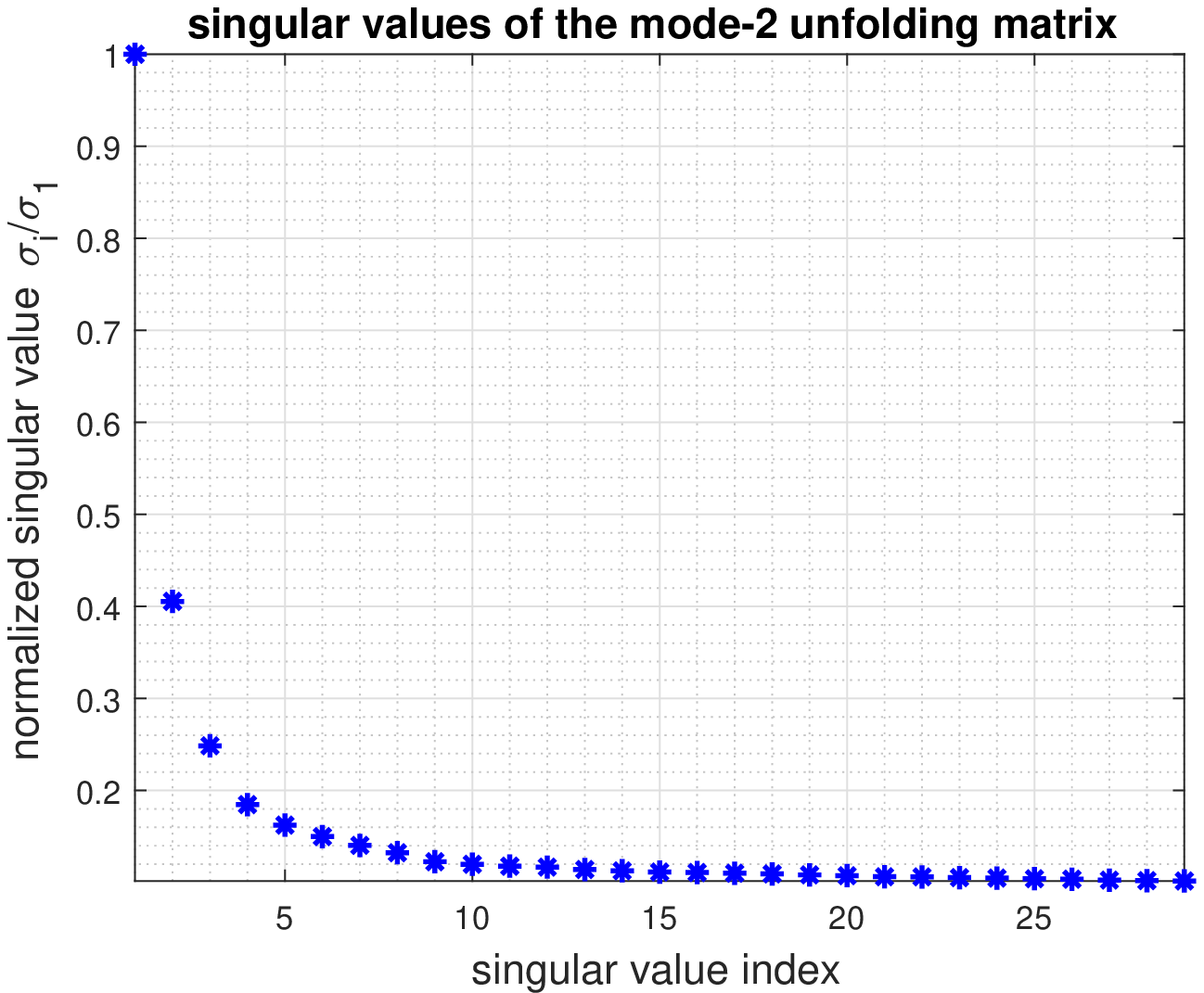}
	~
	\includegraphics[width=0.32\textwidth]{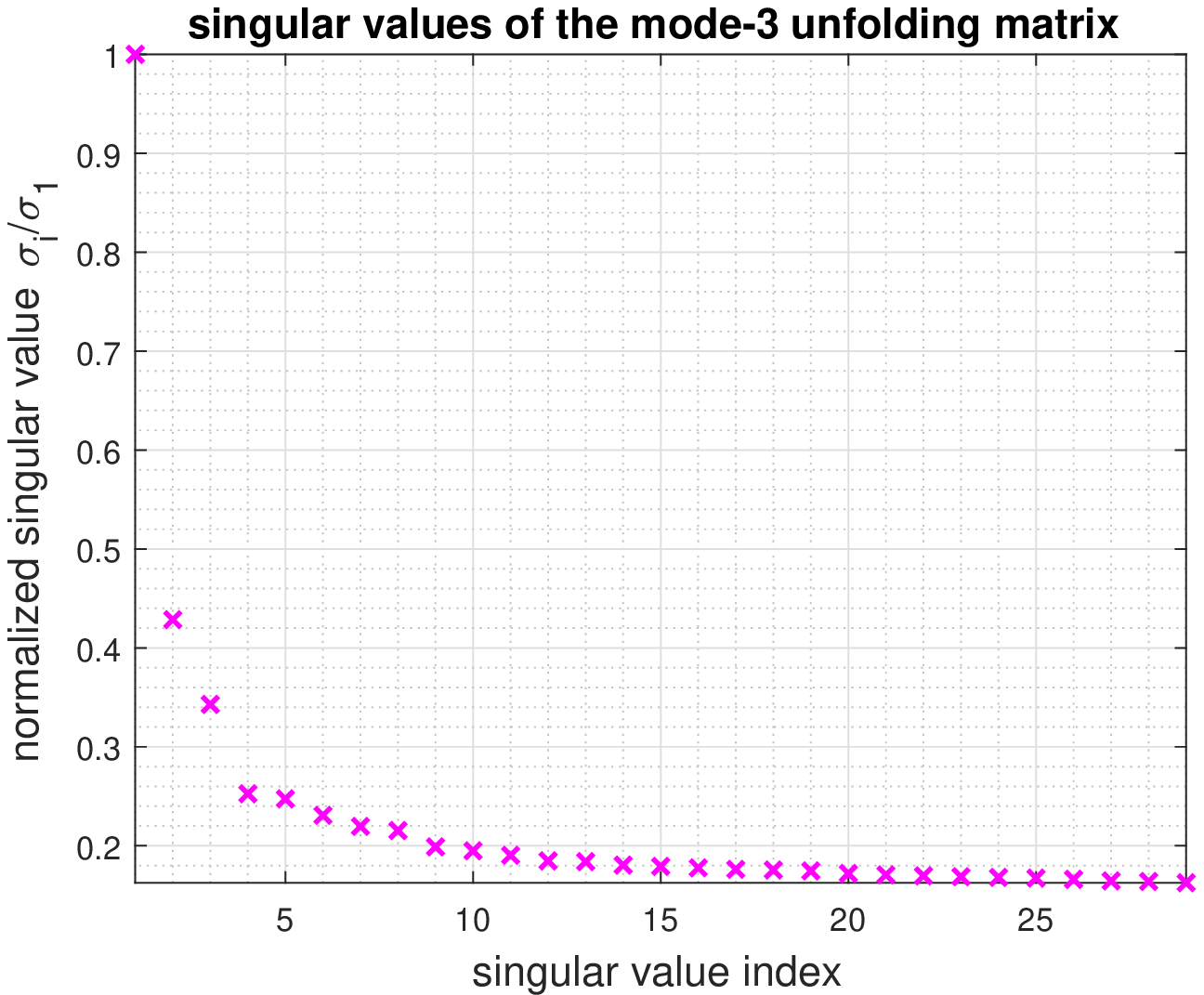}
	\caption{Plots of the normalized singular values of mode-1, -2 and -3 unfolding matrices of the complex-valued InSAR phase stack shown in Figure \ref{fg:roof_area_singular_values}. For visualization, we just plot the first $ 29 $ normalized singular values of mode-1 and -2 unfolding matrices. It is demonstrated that the normalized singular values of the three unfolding matrices decay rapidly, and most of them are below $ 0.2 $, which indicates low rank structures of InSAR phase tensors in real cases.}
	\label{fg:roof_area_3mode_singular_values}

\end{figure*}

In order to investigate the low rank property of an InSAR phase tensor, the normalized singular values ($ \sigma_i/\max(\sigma_i) $) of the mode-1, -2 and -3 unfolding matrices of a simulated noise-free complex-valued phase tensor $ \overline{\mathcal{G}}\in \mathbb{C}^{100\times 100 \times 50} $ (shown in Figure \ref{fg:simu_InSAR_phase_stack}) are plotted in Figure \ref{fg:mode-1,2,3_SV}. It can be observed that the singular values of the three unfolding matrices decay rapidly, which indicates the low rank nature of the original tensor. The low rank tensor approximation $ \tilde{\mathcal{G}} $ of $ \overline{\mathcal{G}} $ can be obtained by the truncated HoSVD with a predefined threshold. As shown in Figure \ref{fg:MSE_compres_ratio_threshold}, we calculate the Mean Square Error (MSE) values of the real-valued residual phases between the approximated tensor $ \tilde{\mathcal{G}} $ and the original tensor $ \overline{\mathcal{G}} $, i.e. $ \mathrm{MSE}(\mathrm{angle}(\tilde{\mathcal{G}}\odot\mathrm{conj}(\overline{\mathcal{G}}))) $, with respect to different thresholds, where $ \odot $ denotes the element-wise product and $ \mathrm{conj}(\cdot) $ is the complex conjugate operator. According to the plot, the original InSAR phase stack can be well approximated by the low rank tensor $ \tilde{\mathcal{G}} $ with acceptable errors. For example, at the thresholding value of $ 0.21 $, the MSE value of the real-valued residual phases between $ \tilde{\mathcal{G}} $ (its multilinear rank is $ (11, 12, 5) $) and $ \overline{\mathcal{G}} $ is around $ 0.01 [\mathrm{rad}^2] $, which is equivalent to an uncertainty of $ 0.2[\mathrm{mm/year}] $ in linear deformation rate or $ 0.69[\mathrm{m}] $ in elevation at the baseline configuration of the simulated data. Such low rank property is often embedded in images. This is especially true in urban areas where man-made objects with regular shapes are abundant.

Such low rank property also exists in real data which usually contains full rank noise. To this end, the normalized singular values of an experimental TerraSAR-X phase tensor with a roof area (Figure \ref{fg:roof_area_singular_values}) are demonstrated in Figure \ref{fg:roof_area_3mode_singular_values}. The associated phase tensor has the dimensions of $ 256\times320\times29 $. It can be seen that the normalized singular values decay rapidly and most of them are below $ 0.2 $, which indicates the low rank structure of the InSAR phase tensor.

\section{Robust phase recovery via iteratively reweighted tensor decomposition}\label{sc:RIRT}
For the case of real data, outliers, e.g. unmodeled phases, usually exist in the observed phase stack. To tackle this challenge, we propose a novel robust tensor decomposition method --- robust iteratively reweighted tensor decomposition.
\subsection{Robust low rank tensor decomposition}
Different from HoSVD where the approximation error is minimized in a least-squares sense, robust low rank tensor decomposition minimizes the rank with $ L_0 $ norm of the approximation error
\begin{equation}
\{\hat{\mathcal{X}},\hat{\mathcal{E}}\}=\argmin_{\mathcal{X},\mathcal{E}}\mathrm{rank}(\mathcal{X})+\gamma\|\mathcal{E}\|_0,\quad s.t.~\mathcal{X}+\mathcal{E}=\mathcal{G},
\label{eq:robust_tensor_recovery}
\end{equation}
where $ \mathcal{G} $ is the observed InSAR phase tensor, $ \mathcal{E} $ models the tensor of sparse outliers, $ \hat{\mathcal{X}},\hat{\mathcal{E}} $ are the recovered outlier-free phase tensor and the estimated outlier tensor, respectively, $ \mathrm{rank}(\mathcal{X}) $ refers to the multilinear rank of $ \mathcal{X} $, $ \|\mathcal{E}\|_0 $ denotes the $ L_0 $ norm of $ \mathcal{E} $, i.e. $ \|\mathcal{E}\|_0=\|\mathrm{vec}(\mathcal{E})\|_0 $, and $ \gamma $ is the regularization parameter.

This problem is NP hard, due to the minimization of the multilinear rank and the $ L_0 $ norm. Regarding this, \cite{goldfarb2014robust} suggested to replace \eqref{eq:robust_tensor_recovery} by the following convex optimization problem
\begin{equation}
\{\hat{\mathcal{X}},\hat{\mathcal{E}}\}=\argmin_{\mathcal{X},\mathcal{E}}\|\mathcal{X}\|_*+\gamma\|\mathcal{E}\|_1,\quad s.t.~\mathcal{X}+\mathcal{E}=\mathcal{G}.
\label{eq:HoRPCA_convex}
\end{equation}
It relaxes the tensor multilinear rank to the tensor nuclear norm $ \|\mathcal{X}\|_* $ which is the sum of the $ N $ nuclear norm $\sum_n\|\mathbf{X}_{(n)}\|_* $ of the mode-$ n $ unfoldings of $ \mathcal{X} $, i.e. $ \|\mathcal{X}\|_*=\sum_n\|\mathbf{X}_{(n)}\|_* $, and by replacing the tensor $ L_0 $ norm with the convex $ L_1 $ norm, i.e. $ \|\mathcal{E}\|_1=\|\mathrm{vec}(\mathcal{E})\|_1 $. This is known as HoRPCA which is a tensor extension of the matrix RPCA \cite{candes2011robust}.

\subsection{Robust Iteratively Reweighted Tensor Decomposition}
In order to better approximate the rank of a matrix and the $ L_0 $ norm of a vector, \cite{candes2008enhancing,peng2014reweighted} proposed a reweighted nuclear norm and $ L_1 $ minimization scheme by enhancing the low rank and sparsity simultaneously during the optimization. The reweighted $ L_1 $ norm is defined as $ \|\mathbf{w}\odot\mathbf{x}\|_1 $, where $ \mathbf{w} $ is the weight vector that updates adaptively for enhancing the sparsity of $ \mathbf{x} $. It is worth noting that if each element of $ \mathbf{w} $ is exactly the inverse absolute value of the corresponding element of $ \mathbf{x} $, i.e. $ w_i=\frac{1}{|x_i|} $, the reweighted $ L_1 $ norm equals the $ L_0 $ norm of $ \mathbf{x} $, i.e. $ \|\frac{1}{|\mathbf{x}|}\odot\mathbf{x}\|_1=\|\mathbf{x}\|_0 $. For the low rank enhancement, the nuclear norm for matrix $ X $ is replaced by a reweighted version $ \|\mathbf{w}\odot\boldsymbol{\sigma}(\mathbf{X})\|_1 $. Likewise, if we have $ w_i=\frac{1}{\sigma_{i}(\mathbf{X})} $, then the rewighted nuclear norm turns into the rank of the matrix $ \mathbf{X} $, i.e. $ \|\mathbf{w}\odot\boldsymbol{\sigma}(\mathbf{X})\|_1=\mathrm{rank}(\mathbf{X}) $.

Inspired by this, we extend the reweighting scheme to the tensor case. By introducing the weights for enhancing the low rank of $ \mathcal{X} $ and the sparsity $ \mathcal{E} $. The optimization problem is
\begin{equation}
\begin{aligned}
\{\hat{\mathcal{X}},\hat{\mathcal{E}}\}=&\argmin_{\mathcal{X},\mathcal{E}}\sum_{n=1}^{N}\|\mathbf{w}_{\mathcal{L},n}\odot\boldsymbol{\sigma}(\mathbf{X}_{(n)})\|_1+\gamma\|\mathcal{W_{\mathcal{E}}}\odot\mathcal{E}\|_1\\& s.t. \quad \mathcal{X}+\mathcal{E}=\mathcal{G},
\end{aligned}
\label{eq:robust_iterative_reweighted_TD}
\end{equation}
where $ \mathbf{w}_{\mathcal{L},n} $ is the weight vector for the singular values of the mode-$ n $ unfolding matrix $ \mathbf{X}_{(n)} $ of $ \mathcal{X} $, and $ \mathcal{W_{\mathcal{E}}} $ is the weight tensor for $ \mathcal{E} $. Note that if all weights are set to $ 1 $, \eqref{eq:robust_iterative_reweighted_TD} will be equivalent to \eqref{eq:HoRPCA_convex}.

\subsection{Optimization by Alternating Direction Method of Multipliers (ADMM)}
The optimization problem \eqref{eq:robust_iterative_reweighted_TD} can be solved by the ADMM framework \cite{boyd2011distributed}. The constraint optimization
problem in \eqref{eq:robust_iterative_reweighted_TD} is firstly converted to its augmented Lagrangian function, yielding
\begin{equation}
\begin{aligned}
L_{\mu}(\mathcal{X},\mathcal{E},\mathcal{Y})=&\sum_{n=1}^{N}\|\mathbf{w}_{\mathcal{L},n}\odot\boldsymbol{\sigma}(\mathbf{X}_{(n)})\|_1+\gamma\|\mathcal{W_{\mathcal{E}}}\odot\mathcal{E}\|_1-\\ &\langle\mathcal{Y},\mathcal{X}+\mathcal{E}-\mathcal{G}\rangle+\frac{1}{2\mu}\|\mathcal{X}+\mathcal{E}-\mathcal{G}\|_F^2,
\end{aligned}
\end{equation}
where $ \mathcal{Y} $ denotes the introduced dual variable and $ \mu $ is the penalty parameter. ADMM takes advantage of splitting one difficult optimization problem into several subproblems, where each of them has a closed-form solution. Accordingly, the minimization of $ L_{\mu} $ with respect to each variable can be solved by optimizing the following subproblems:

\subsubsection{$ \mathcal{X} $ subproblem}
By fixing $ \mathcal{E} $ and $ \mathcal{Y} $, the subproblem of $ L_{\mu} $ with respect to $ \mathcal{X} $ can be rewritten as
\begin{equation}
\min_{\mathcal{X}}\sum_{n=1}^{N}\|\mathbf{w}_{\mathcal{L},n}\odot\boldsymbol{\sigma}(\mathbf{X}_{(n)})\|_1+\frac{1}{2\mu}\|\mathcal{X}+\mathcal{E}-\mathcal{G}-\mu\mathcal{Y}\|_F^2.
\label{eq:X_subproblem}
\end{equation}
This subproblem can be solved by the Nonuniform Singular Value Thresholding (NSVT) operator \cite{peng2014reweighted,gu2014weighted}. Taking matrix $ \mathbf{A} $ as an example, given the thresholding weight vector $ \mathbf{w} $, NSVT is defined as $ \mathcal{T}_{\mathbf{w}}(\mathbf{A}):=\mathbf{U}\mathrm{diag}(\max(\sigma_{i}-w_i,0))\mathbf{V} $, with $ \mathbf{U} $, $ \mathbf{V} $ and $ \sigma_{i} $ calculated by SVD of $ \mathbf{A} $.
\subsubsection{$ \mathcal{E} $ subproblem}
By fixing $ \mathcal{X} $ and $ \mathcal{Y} $, the subproblem of $ L_{\mu} $ with respect to $ \mathcal{E} $ has the following form
\begin{equation}
\min_{\mathcal{E}}\gamma\|\mathcal{W_{\mathcal{E}}}\odot\mathcal{E}\|_1+\frac{1}{2\mu}\|\mathcal{X}+\mathcal{E}-\mathcal{G}-\mu\mathcal{Y}\|_F^2.
\label{eq:E_subproblem}
\end{equation}
This weighted $ L_1 $-norm optimization subproblem can be solved by the Nonuniform Soft Thresholding (NST) operator, which is defined as $ \mathcal{S}_{\mathcal{W}}(\mathcal{A}):=\mathrm{sign}(\mathcal{A})\odot\max(|\mathcal{A}|-\mathcal{W},0) $, with $ |\mathcal{A}|=\mathrm{sign}(\mathcal{A})\odot\mathcal{A}$.
\subsubsection{$ \mathcal{Y} $ updating}
The dual variable $ \mathcal{Y} $ can be updated by
\begin{equation}
\mathcal{Y}=\mathcal{Y}-\frac{1}{\mu}(\mathcal{X}+\mathcal{E}-\mathcal{G}).
\label{eq:Y_update}
\end{equation}
\subsubsection{Weight updating}
The weight vector $ \mathbf{w}_{\mathcal{L},n}, n=1,\ldots,N $ and the weight tensor $ \mathcal{W}_{\mathcal{E}} $ can be updated by
\begin{equation}
	\mathbf{w}_{\mathcal{L},n} = \frac{1}{\boldsymbol{\sigma}(\mathbf{X}_{(n)})+\epsilon_{\mathcal{L}}},\quad\quad\mathcal{W}_{\mathcal{E}}=\frac{1}{|\mathcal{E}|+\epsilon_{\mathcal{E}}},
	\label{eq:weights_update}
\end{equation}
where $ \epsilon_{\mathcal{L}} $ and $ \epsilon_{\mathcal{E}} $ are the predetermined positive constants.

The detailed ADMM pseudocode for solving \eqref{eq:robust_iterative_reweighted_TD} is summarized in Algorithm \ref{ag:2}.

Using a predefined convergence condition, the solution $ (\hat{\mathcal{X}}, \hat{\mathcal{E}}) $ can be obtained, i.e. the outlier-free InSAR phase tensor and the sparse outlier tensor, respectively. To this end, by applying conventional multipass InSAR techniques, e.g. PSI \cite{ferretti2001permanent}, on $ \hat{\mathcal{X}} $, we can robustly retrieve the geophysical parameters.

\begin{algorithm}
	\caption{RoMIO solved by ADMM}
	\begin{algorithmic}[1]
		\Require $ \mathcal{G},\gamma,\mu,N,\epsilon_{\mathcal{L}}=\epsilon_{\mathcal{E}}=1\times10^{-3} $ 
		\State Initialize $ \mathcal{X}^{(0)}=\mathcal{E}^{(0)}=\mathcal{Y}^{(0)}=0 $
		\For {$ k=0 $ to $ k_{\max} $}
		\State NSVT on the mode-$ n, n=1,\ldots,N $ unfolding of $ \mathcal{G}+\mu\mathcal{Y}^{(k)}-\mathcal{E}^{(k)} $, \par
		then, folding mode-$ n $ tensors and averaging them by $ N $\par
		$ \mathcal{X}^{(k+1)}\leftarrow\frac{1}{N}\sum_{n=1}^{N}\mathcal{T}_{n,\mu N\mathbf{w}^{(k)}_{\mathcal{L},n}}(\mathbf{G}_{(n)}+\mu\mathbf{Y}_{(n)}^{(k)}-\mathbf{E}_{(n)}^{(k)}) $, \par
		where $ \mathcal{T}_{n,\mu N\mathbf{w}^{(k)}_{\mathcal{L},n}}(\cdot):=\mathrm{fold}_n(\mathcal{T}_{\mu N\mathbf{w}^{(k)}_{\mathcal{L},n}}(\cdot)) $,
		\State NST on the the tensor $ \mathcal{G}+\mu\mathcal{Y}^{(k)}-\mathcal{X}^{(k+1)} $: \par
		$ \mathcal{E}^{(k+1)}\leftarrow\mathcal{S}_{\mu\gamma\mathcal{W}^{(k)}_{\mathcal{E}}}(\mathcal{G}+\mu\mathcal{Y}^{(k)}-\mathcal{X}^{(k+1)}) $,
		\State $ \mathcal{Y}^{(k+1)}\leftarrow\mathcal{Y}^{(k)}-\frac{1}{\mu}(\mathcal{X}^{(k+1)}+\mathcal{E}^{(k+1)}-\mathcal{G}) $,
		\State Updating weights: \par
		$ \mathbf{w}^{(k+1)}_{\mathcal{L},n} = \frac{1}{\boldsymbol{\sigma}(\mathbf{X}^{(k+1)}_{(n)})+\epsilon_{\mathcal{L}}},\quad\mathcal{W}^{(k+1)}_{\mathcal{E}}=\frac{1}{|\mathcal{E}^{(k+1)}|+\epsilon_{\mathcal{E}}} $,
		\If{convergence}
		\State	break
		\EndIf
		\EndFor
		\Ensure $ (\hat{\mathcal{X}},\hat{\mathcal{E}}) $
	\end{algorithmic}
\label{ag:2}
\end{algorithm}

\begin{figure*}
	\centering
	\includegraphics[width=\textwidth]{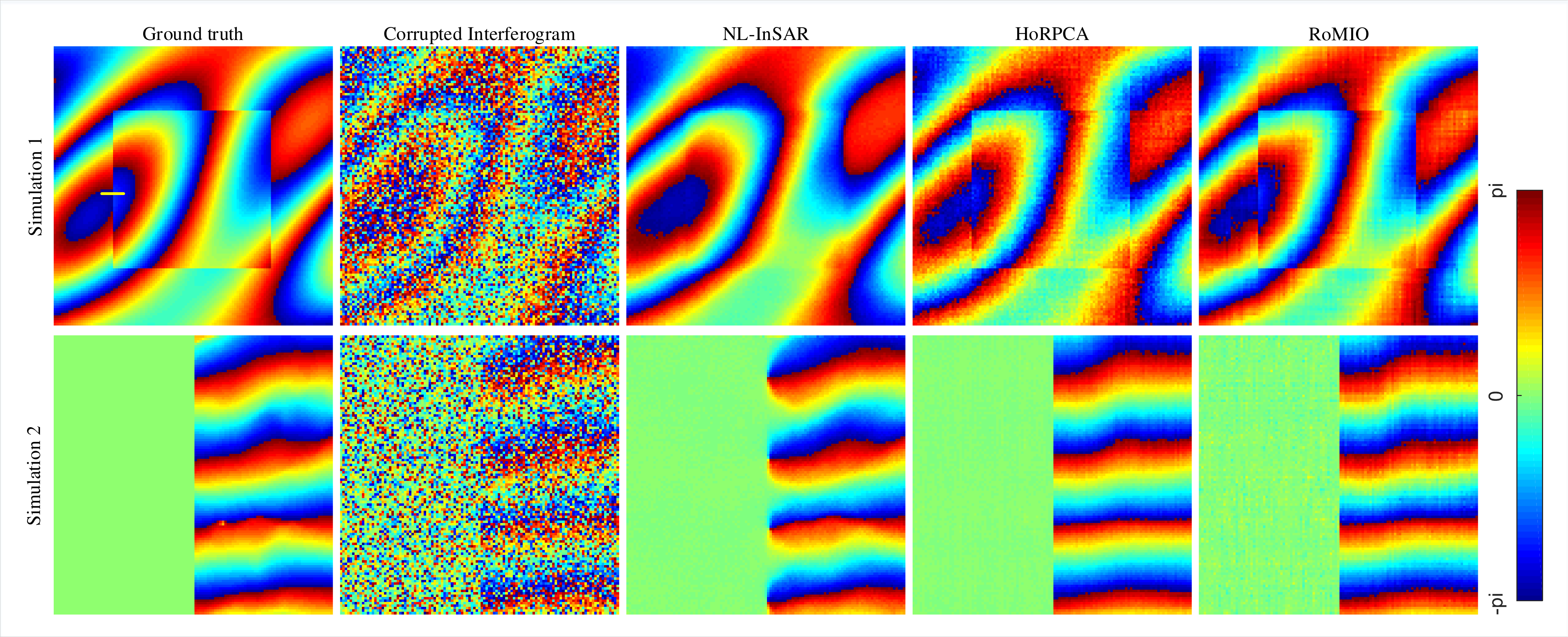}
	\caption{Plots of one interferogram in the two simulated InSAR phase stacks, generated by the corresponding geophysical parameters shown in Figure \ref{fg:simu_mountain_realmountain_DeforEle_results}, as well as the corrupted phases with an SNR of 5dB and $ 30\% $ outliers, and the recovered results by three methods. Although the NL-InSAR result can maintain the smooth fringes very well, the edges of rectangle in the middle are more blurred compared to the other two results. This can be clearly observed at the two cropped parts in Figure \ref{fg:one_profile_simu_inter_phase}. Compared to HoRPCA, the proposed method can better keep the original structure of the interferogram, since it can better capture the low rank structure of the data and model the sparse outliers by enhancing the low rank and the sparsity.}
	\label{fg:simu_interferograms_NL_HoRPCA_RIRTD}
\end{figure*}
\begin{figure}
	\centering
	\includegraphics[width=0.5\textwidth]{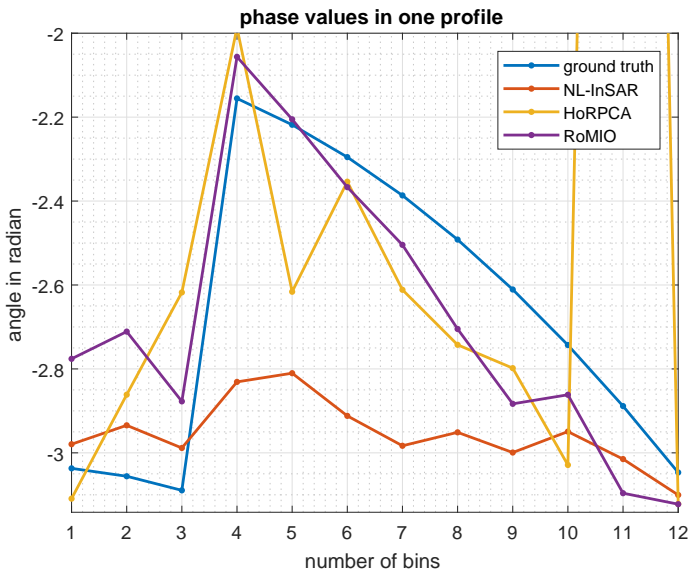}
	\caption{Profiles of the estimated phases marked by the short yellow line segment in Figure \ref{fg:simu_interferograms_NL_HoRPCA_RIRTD}. It is obvious to show that the estimations of this area are blurred in the NL-InSAR result compared with the others.}
	\label{fg:one_profile_simu_inter_phase}
\end{figure}
\begin{figure}
	\centering
	\includegraphics[width=0.45\textwidth]{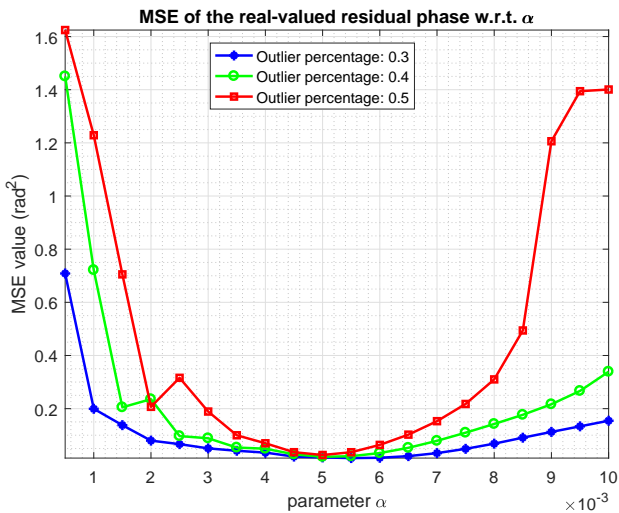}
	\caption{Plot of the MSE values of the real-valued residual phases between the phase tensor (Simulation 1) recovered by RoMIO and its ground truth, with respect to different parameter ($ \alpha $) values. As shown in the figure, even under a high percentage of outliers, e.g. $ 30\% $, the operative range of $ \alpha $ still keeps relatively wide. Of course, this range decreases as the percentage of outliers increases. Also, the parameter can also be tuned using the L-curve method \cite{hansen1993use,kang2017robust}. Still, for a particular dataset, the optimal $ \alpha $ for different percentages of outliers is similar (around $ 5\times10^{-3} $ in our simulation), which means that no assumptions about the amount of outliers is required.}
	\label{fg:RIRTD_para_setting}
\end{figure}
\begin{figure*}
	\centering
	\includegraphics[width=\textwidth]{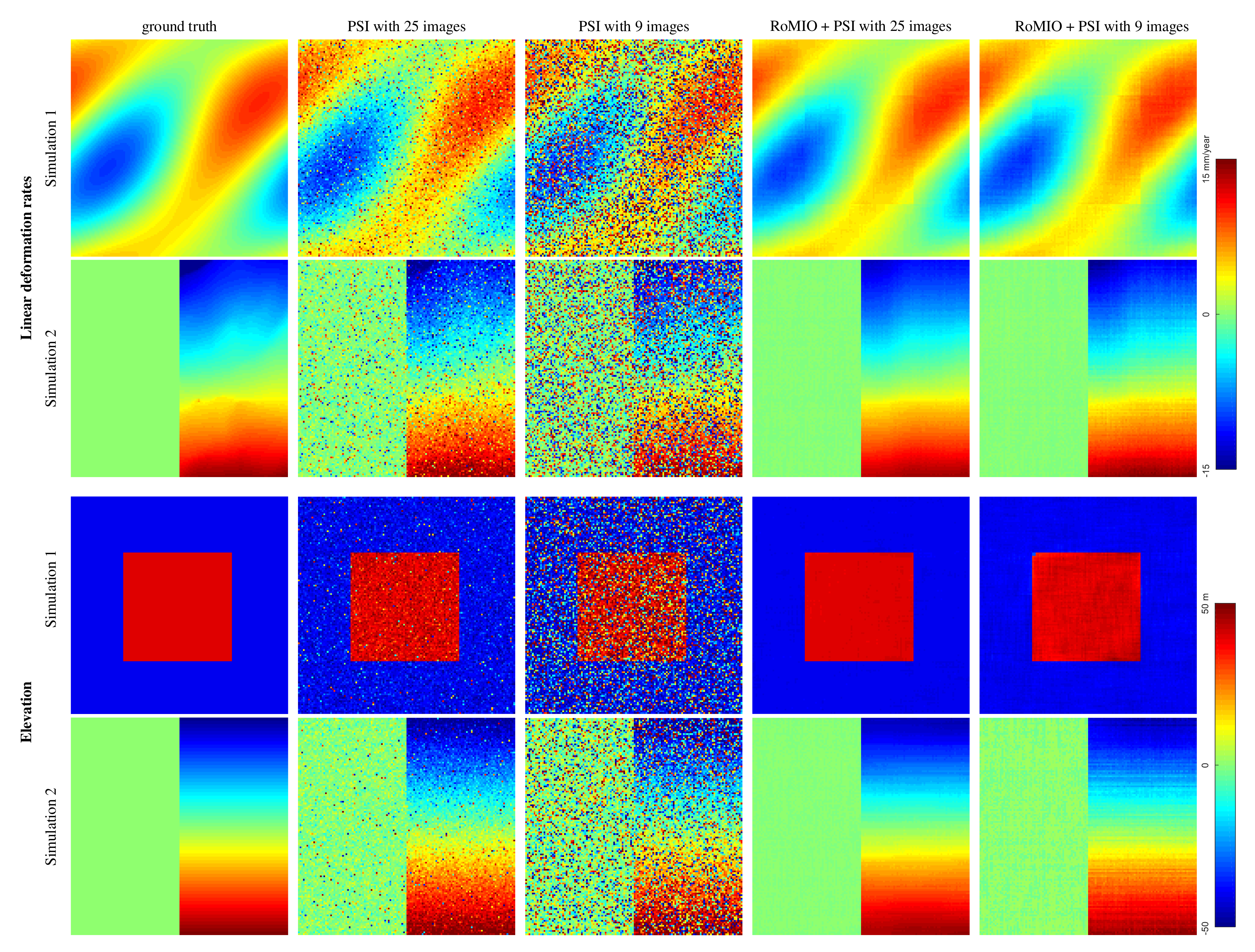}
	\caption{The simulated ground truth linear deformation rates and elevations of the two simulations, along with the estimated results by PSI and RoMIO + PSI with 25 and 9 SAR images. The results of PSI contain outliers. This is especially true for the result from a subset of the stack. The reason is that periodogram method in PSI is only asymptotically optimal, which means large bias is very likely to occur at low number of images. In contrast, the proposed method can robustly recover the parameters both using the full stack and a subset of the stack. That is to say the proposed method can in turn effectively reduce the number of images required for a reliable estimation.}
	\label{fg:simu_mountain_realmountain_DeforEle_results}
\end{figure*}

\begin{figure}
	\centering
	\includegraphics[width=0.5\textwidth]{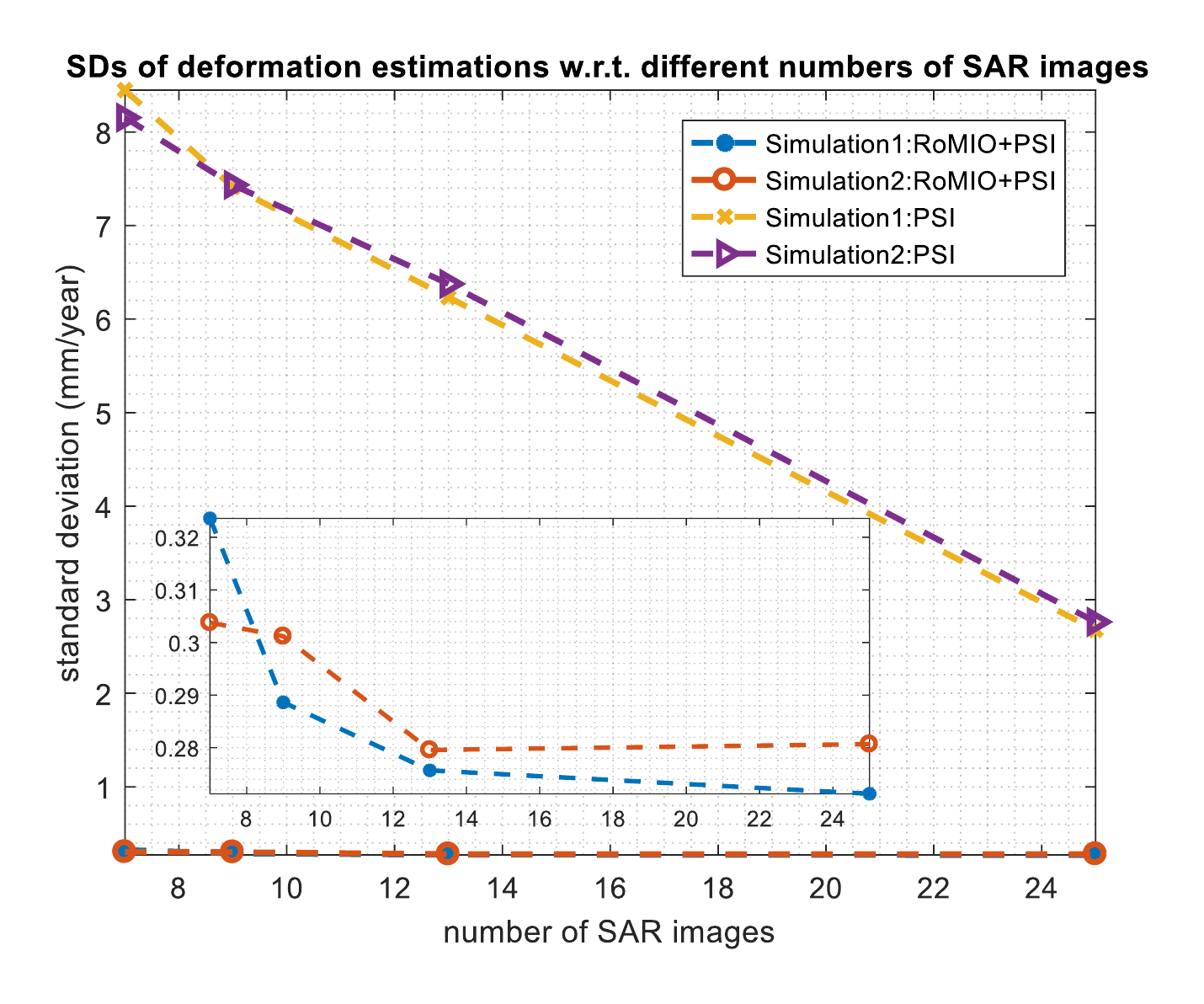}
	\caption{Plot of SDs of deformation estimations with respect to different numbers of SAR images for reconstruction. The proposed method can achieve a SD around $ 0.3[\mathrm{mm/year}] $, which can improve the estimation accuracy of PSI more than ten times. It shows the accuracy of RoMIO + PSI can maintain at a better and more constant level compared to the PSI whose efficiency decreases linearly w.r.t. the number of images. At the number of  images down to $ 7 $, the accuracy of RoMIO + PSI still keeps at a sub millimeter range which is about 30 times better than PSI. This creates an opportunity of multipass InSAR geophysical parameter reconstruction using very small stacks.}
	\label{fg:SD_defor_dif_num_SAR_img}

\end{figure}

%

\section{Experimental results}\label{sc:experiments}
\subsection{Simulations}
We simulated two multipass InSAR phase stacks of $ 128\times128 $ pixels by $ 25 $ images with different spatial patterns. The corresponding linear deformation and elevation maps are shown in Figure \ref{fg:simu_mountain_realmountain_DeforEle_results}. Note that the two geophysical maps of Simulation 1 are spatially uncorrelated, while those of Simulation 2 are highly correlated. Their linear deformation rates both range from $ -15 [\mathrm{mm/year}] $ to $ 15 [\mathrm{mm/year}] $ and elevation values are from $ -50 [\mathrm{m}] $ to $ 50 [\mathrm{m}] $. The spatial baseline and the temporal baseline were chosen to be comparable to those of TerraSAR-X.  Uncorrelated complex circular Gaussian noise was added to the two simulated stacks with an SNR of 5dB, i.e. according to PS model. To simulate sparse outliers in the stacks, $ 30\% $ of pixels randomly selected from the phase tensor were replaced with uniformly distributed phases.

For visualizing the performance of the proposed method, we chose one interferogram from the recovered phase tensor $ \hat{\mathcal{X}} $ and visually compared it with those obtained by NL-InSAR \cite{deledalle2011nl} and HoRPCA \cite{kang2017robust} in Figure \ref{fg:simu_interferograms_NL_HoRPCA_RIRTD}. Since NL-InSAR is designed for denoising one interferogram, whereas the others make use of the full image stack, to achieve a relatively fair comparison, the NL-InSAR result was obtained by averaging the results from $ 25 $ simulations of InSAR phase stacks. In our method, the spatial size of the tensor is set as $ 128\times128 $ (i.e. the whole stack as one tensor), $ \gamma $ is set to be $ 4.4\times 10^{-4} $ and $ \mu $ is kept constant at the value $ 10\times\mathrm{std}(\mathrm{vec}(\mathcal{G})) $. The experiments for the associated parameter setting will be introduced in the following section. The search window size and the patch size in NL-InSAR is $ 21\times21 $ and $ 5\times5 $, respectively. In addition, the phase profile marked by the short yellow line segment in Figure \ref{fg:simu_interferograms_NL_HoRPCA_RIRTD}, are plotted in Figure \ref{fg:one_profile_simu_inter_phase}. For a quantitative evaluation, we list the MSE values of the real-valued residual phases between the recovered phase tensor and the ground truth, i.e. $ \mathrm{MSE}(\mathrm{angle}(\hat{\mathcal{X}}\odot\mathrm{conj}(\mathcal{X}))) $, in cases of $ 30\% $, $ 40\% $ and $ 50\% $ percentages of outliers in Table \ref{tb:MSE_performance}.

\begin{table*}
	\caption{MSE performances of NL-InSAR, HoRPCA and RoMIO on the simulations shown in Figure \ref{fg:simu_interferograms_NL_HoRPCA_RIRTD}}
	\centering
	\begin{tabular}{ c| c| c| c| c| c| c| c }
		\hline
		\hline
		\multicolumn{2}{ c |}{} & \multicolumn{3}{ c |}{Simulation 1} & \multicolumn{3}{ c  }{Simulation 2} \\
		\multicolumn{2}{ c |}{} & NL-InSAR & HoRPCA  & RoMIO & NL-InSAR & HoRPCA  & RoMIO \\
		\hline
		\multirow{3}{*}{\parbox{3cm}{Mean Square Error (MSE) [$ \mathrm{rad}^2 $]}} & $ 30\% $ outliers & $ \mathbf{0.03} $ & $ 0.04 $ & $ \mathbf{0.03} $ & $ \mathbf{0.03} $ & $ 0.04 $ & $ 0.04 $ \\
		\cline{2-8}
		& $ 40\% $ outliers & $ 0.05 $ & $ 0.07 $ & $ \mathbf{0.04} $ & $ \mathbf{0.05} $ & $ 0.06 $ & $ \mathbf{0.05} $ \\
		\cline{2-8}
		& $ 50\% $ outliers & $ 0.12 $ & $ 0.12 $ & $ \mathbf{0.06} $ & $ 0.07 $ & $ 0.12 $ & $ \mathbf{0.06} $ \\
		\hline
		\hline
	\end{tabular}
	\label{tb:MSE_performance}
\end{table*}

Furthermore, we compared the estimated results of geophysical parameters by PSI and the proposed RoMIO + PSI, using the simulated data. The outlier percentage was set to $ 30\% $ and SNR was 5dB. $ \alpha $ was set to $ 5\times10^{-3} $. The results are illustrated in Figure \ref{fg:simu_mountain_realmountain_DeforEle_results}. The first two rows are the estimates of linear deformation rates of the two simulations and the last two rows are the corresponding elevation estimates. In addition to the experiments based on the full stack of $ 25 $ SAR images, experiments using only $ 9 $ images were conducted in order to test the RoMIO's capability to handle small stacks. For the associated quantitative evaluation, we calculated both bias and standard deviation (SD) of the results and present them in Table \ref{tb:SD_study_Defor_Ele}. To study the minimum number of images for RoMIO to achieve a reliable estimation, we plot the SD of the deformation estimates obtained by RoMIO + PSI w.r.t a decreasing number of SAR images down to $ 7 $.


\begin{table*}
	\centering
	\caption{Quantitative study of the results in Figure \ref{fg:simu_mountain_realmountain_DeforEle_results}}
	\label{tb:SD_study_Defor_Ele}
	\begin{tabular}{ c| c| c| c| c| c}
		\hline
		\hline
		\multicolumn{2}{c|}{} & \multicolumn{2}{c|}{Deformation [mm/year]} & \multicolumn{2}{c}{Elevation [m]} \\
		\cline{3-6}
		\multicolumn{2}{c|}{} & SD & bias & SD & bias \\
		\cline{3-6}
		\hline
		\multirow{4}{*}{simulation 1} & PSI (25 images) & $ 2.68 $ & $ -0.01 $ & $ 8.18 $ & $ 0.42 $ \\
		\cline{2-6}
		& PSI (9 images) &  $ 7.41 $ & $ -0.04 $ & $ 31.56 $ & $ 7.02 $\\
		\cline{2-6}
		& RoMIO+PSI (25 images)  & $ 0.27 $ & $ 0.01 $ & $ 0.39 $ & $ -0.05 $\\
		\cline{2-6}
		& RoMIO+PSI (9 images)  & $ 0.29 $ & $ -0.02 $ & $ 1.59 $ & $ 0.01 $\\
		\hline
		\multirow{4}{*}{simulation 2} & PSI (25 images) & $ 2.76 $ & $ 0.02 $ & $ 7.27 $ & $ 0.07 $ \\
		\cline{2-6}
		& PSI (9 images) & $ 9.16 $ & $ 0.05 $ & $ 21.12 $ & $ 0.21 $ \\
		\cline{2-6}
		& RoMIO+PSI (25 images)  & $ 0.31 $ & $ 0.01 $ & $ 0.98 $ & $ 0.02 $ \\
		\cline{2-6}
		& RoMIO+PSI (9 images)  & $ 0.31 $ & $ -0.01 $ & $ 1.17 $ & $ 0.13 $ \\
		\hline
		\hline
	\end{tabular}
\end{table*}
\subsection{Real data}
\subsubsection{Berlin bridge}
\begin{figure*}
	\centering
	\includegraphics[width=\textwidth]{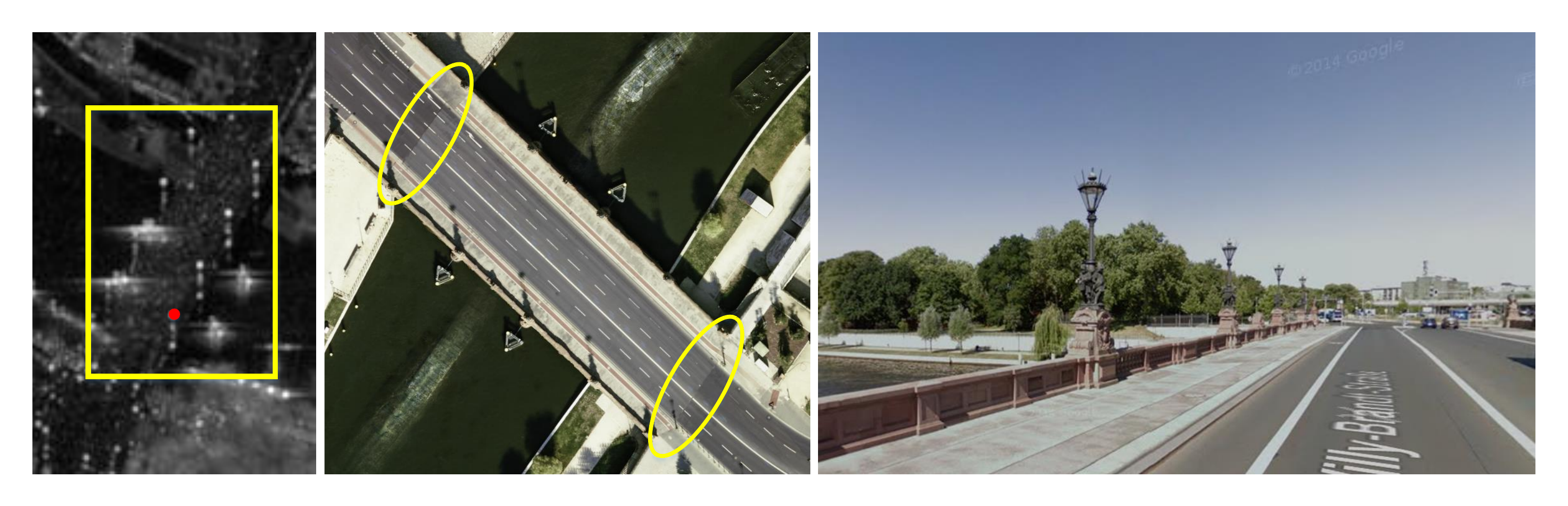}
	\caption{(Left) The TerraSAR-X test image of one bridge area in Berlin as cropped by the yellow rectangle. The red point is the reference point for the elevation and seasonal motion reconstruction in this area. (Middle) The associated orthorectified optical image, generated using semi-global matching \cite{hirschmuller2008stereo}. (Right) The streetview image from Google StreetView.}
	\label{fg:berlin_bridge_SAR_optical_streetview}
\end{figure*}

\begin{figure}
	\centering
	\includegraphics[width=0.5\textwidth]{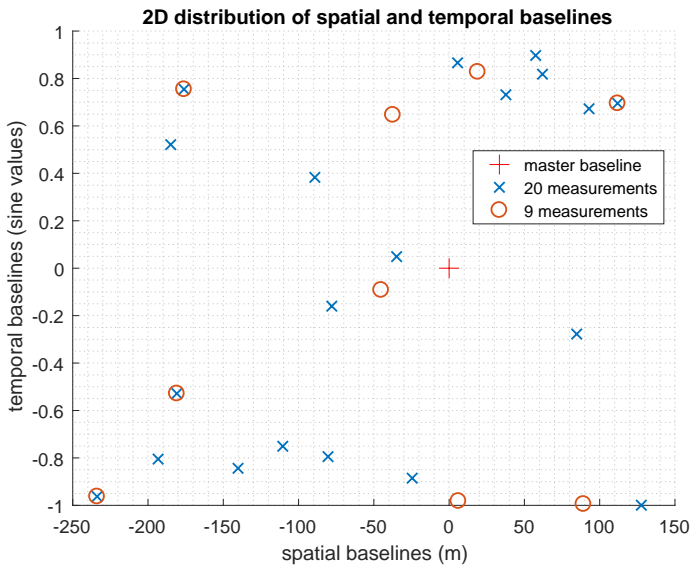}
	\caption{The 2D distribution of spatial and temporal baselines of the selected $ 20 $ and $ 9 $ measurements for reconstruction. The baselines were also chosen to be close to uniform distribution.}
	\label{fg:berlin_bridge_baselines}
\end{figure}

The first TerraSAR-X test area is a bridge in Berlin which is marked by the yellow rectangle shown in Figure \ref{fg:berlin_bridge_SAR_optical_streetview} (Left), where the reference point for the elevation and seasonal motion reconstruction is plotted in red. To its right, the corresponding orthorectified optical image \cite{hirschmuller2008stereo} and a streetview image from Google StreetView are also displayed. The InSAR stack contains over a hundred images. However, in order to test the performance under low number of images, $ 20 $ and $ 9 $ SAR images were selected from the full stack, respectively. They were selected to be similar in their distributions and spans of the temporal and spatial baselines, so that the Cram{\'e}r-Rao bounds of the estimates are comparable. The baselines were also chosen to be close to uniform distribution. The 2D baseline distribution of the selected images can be seen in Figure \ref{fg:berlin_bridge_baselines}. The estimated amplitudes of the seasonal motion and the elevation by PSI and RoMIO + PSI are demonstrated in Figure \ref{fg:geo_para_berlin_brdige_results}.

\begin{figure}
	\centering
	\includegraphics[width=0.5\textwidth]{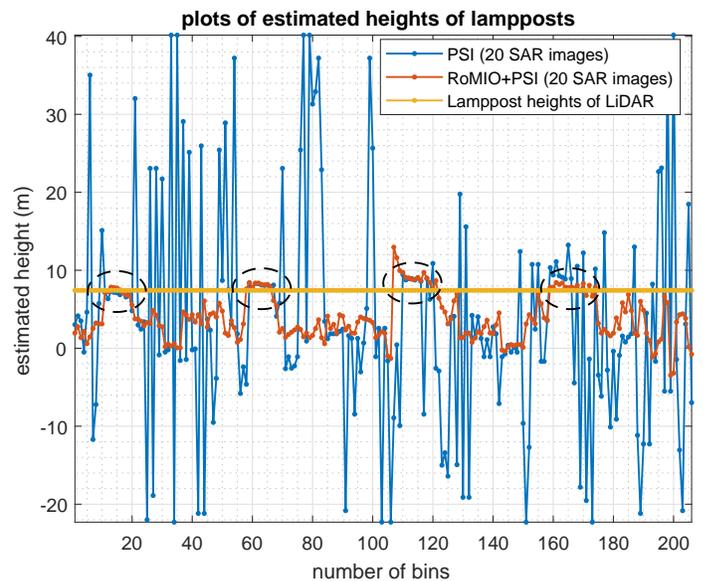}
	\caption{The extracted two profiles of height estimates located at the yellow arrow positions of the results of PSI and RoMIO + PSI, along with the lamppost height profile of LiDAR. Obviously, the four lampposts (shown by the black dash ellipses) are well distinguishable in the result of the proposed method.}
	\label{fg:berlin_bridge_ele_profiles}
\end{figure}
\begin{figure*}
	\centering
	\includegraphics[width=\textwidth]{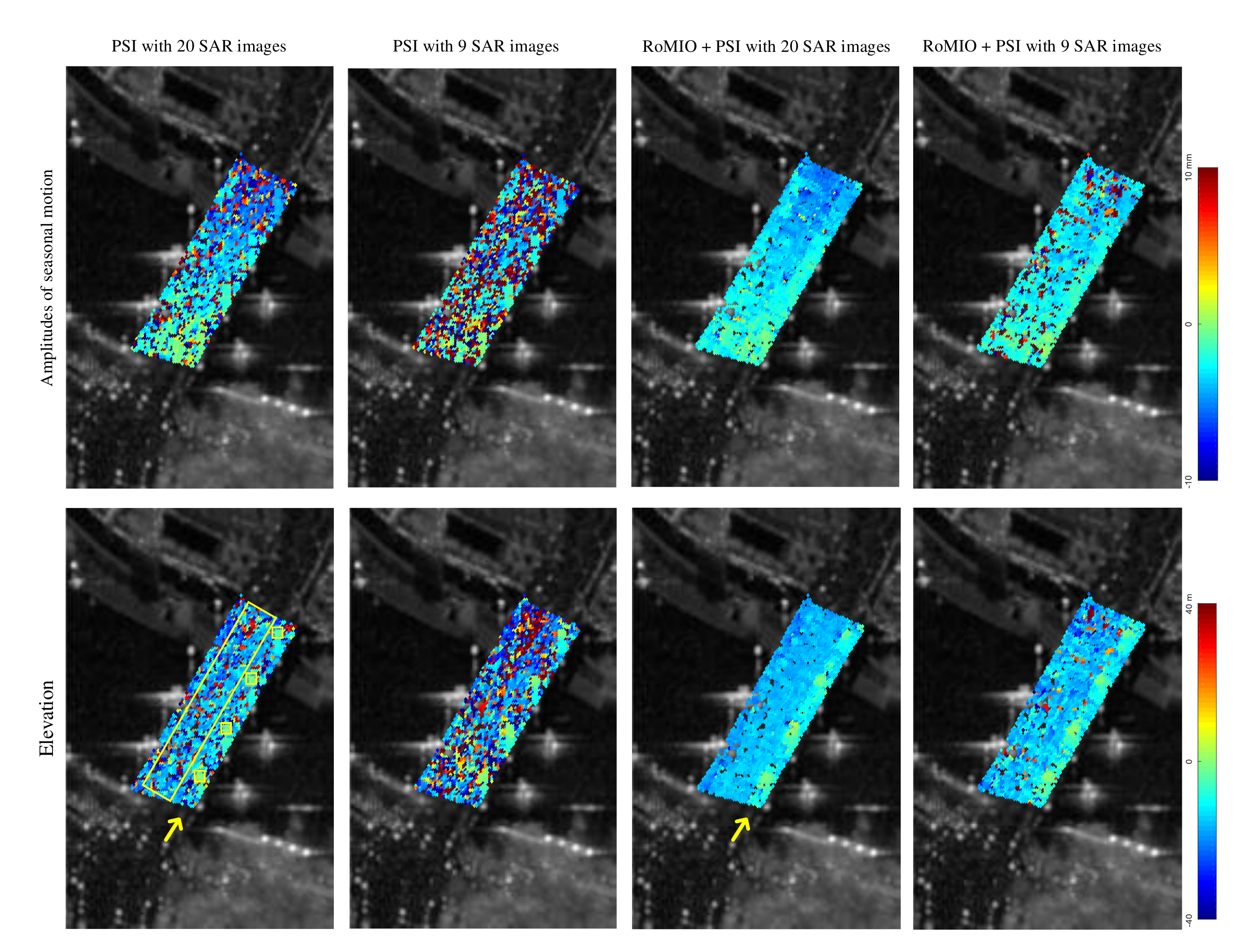}
	\caption{Geophysical parameter estimations (amplitudes of the seasonal motion and elevations) of the area by PSI and RoMIO + PSI with 20 and 9 SAR images. Consistent with the simulations, the proposed method can achieve a more robust estimation result than the classical PSI. In particular, under limited number of images, the interpretation of the parameters retrieved by PSI is severely influenced by outliers. The results of the proposed method are more interpretable. One can observe that the amplitudes of the motion tend to increase from one sider to the other. One plausible reason is that the deformation allowances on the two sides of the bridge are different. To verify this, a very high resolution image of the bridge is shown in Figure \ref{fg:berlin_bridge_SAR_optical_streetview} (Middle). Interesting to note is that there are four elevated regions which correspond to the four lampposts on the bridge. We plot the corresponding two profiles from the results of PSI and RoMIO + PSI in Figure \ref{fg:berlin_bridge_ele_profiles}.}
	\label{fg:geo_para_berlin_brdige_results}
\end{figure*}
\begin{figure}
	\centering
	\includegraphics[width=0.45\textwidth]{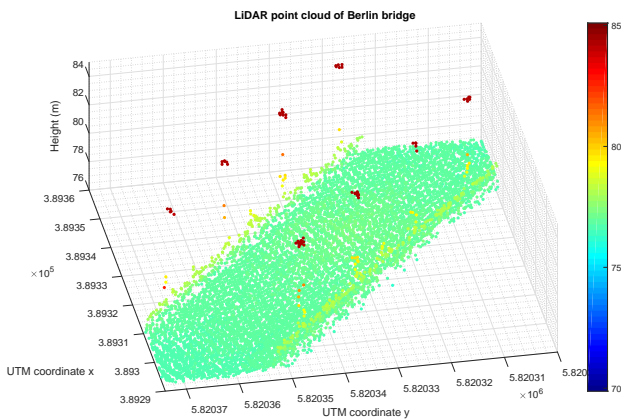}
	\caption{LiDAR point cloud of the study Berlin bridge}
	\label{fg:lidar_berlin_bridge}
\end{figure}

\begin{table*}
	\centering
	\caption{Lamppost height estimations of the two methods with 20 SAR images, along with the reference of LiDAR point cloud.}
	\label{fg:LiDAR_Berlin_bridge_lamppost}
	\begin{tabular}{c | c| c | c | c}
		\hline
		\hline
		Unit (m) & Lamppost1 & Lamppost2 & Lamppost3 & Lamppost4 \\
		\hline
		LiDAR height & \multicolumn{4}{c}{$ 7.42 $} \\
		\hline
		PSI mean & $ 6.76 $ & $ 7.70 $ & $ 8.82 $ & $ 10.03 $ \\
		\hline
		PSI SD  & $ 1.39 $ & $ 2.17 $ & $ 0.26 $ & $ 2.56 $ \\
		\hline
		RoMIO+PSI mean & $ 7.13 $ & $ 7.69 $ & $ 8.91 $ & $ 8.01 $ \\
		\hline
		RoMIO+PSI SD  & $ 1.30 $ & $ 1.74 $ & $ 0.28 $ & $ 0.46 $ \\
		\hline
		\hline
	\end{tabular}

\end{table*}

\subsubsection{Las Vegas convention center}
\begin{figure*}
	\centering
	\includegraphics[width=\textwidth]{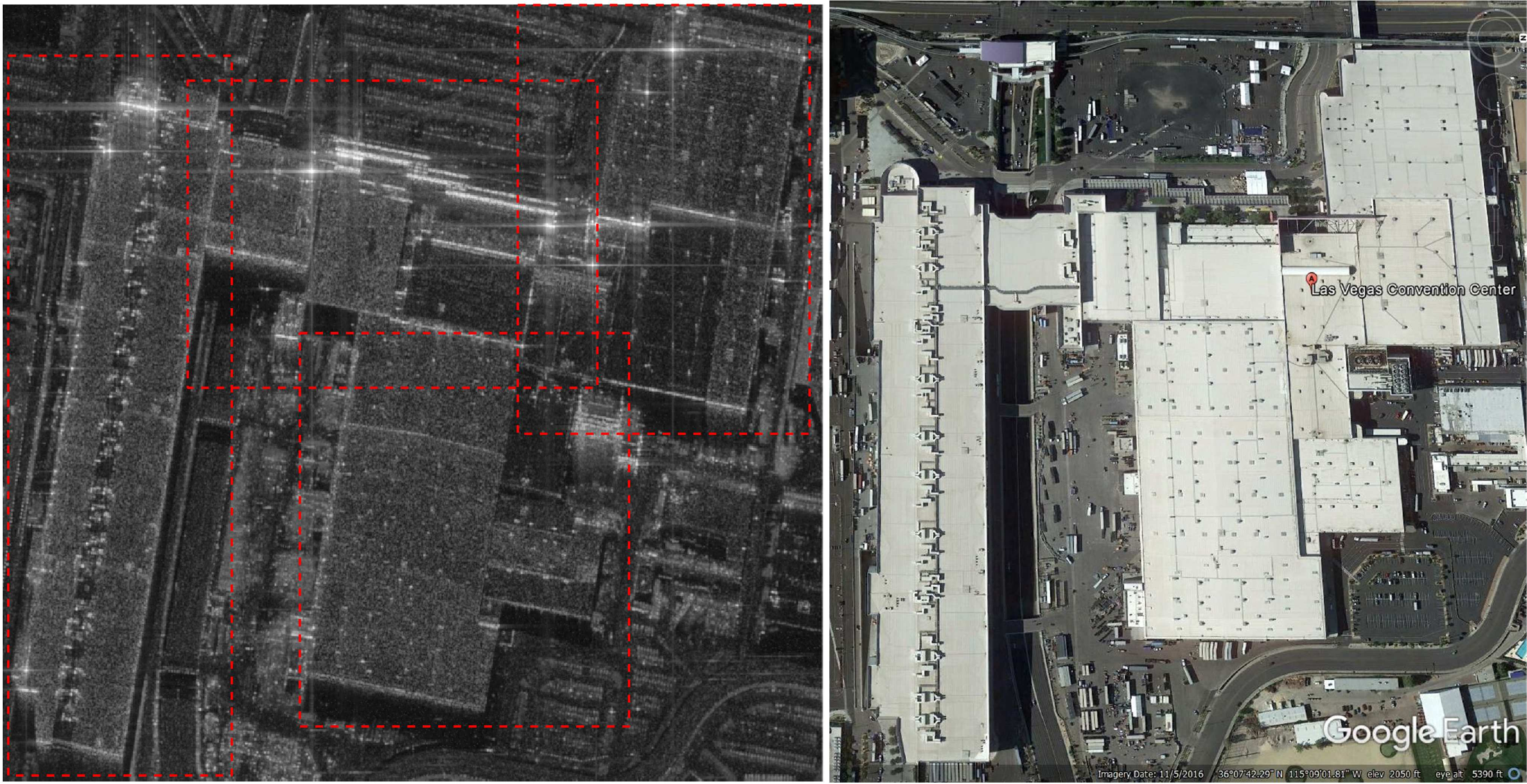}
	\caption{(Left) The TerraSAR-X test image of the Las Vegas convention center. Since the building structure is complex and its spatial area is large ($ 800\times850 $ pixels), we separately process the four parts of the whole InSAR phase stack as cropped with the red dashed rectangles in the figure. (Right) The associated optical image from Google Earth.}
	\label{fg:lasvegas_roof_SAR_google}
\end{figure*}

\begin{figure}
	\centering
	\includegraphics[width=0.5\textwidth]{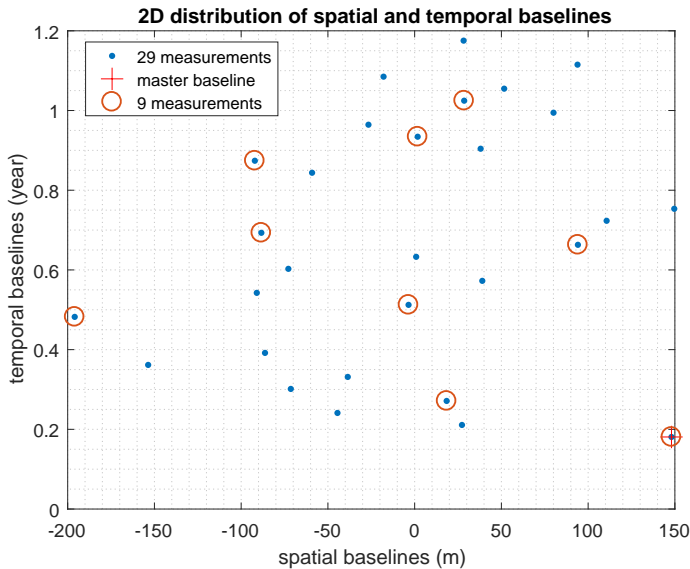}
	\caption{The 2D distribution of spatial and temporal baselines of the total $ 29 $ measurements is demonstrated, along with those of the selected $ 9 $ measurements for reconstruction. The baselines were also chosen to be close to uniform distribution.}
	\label{fg:lasvegas_2D_baselines}
\end{figure}

\begin{figure*}
	\centering
	\includegraphics[width=\textwidth]{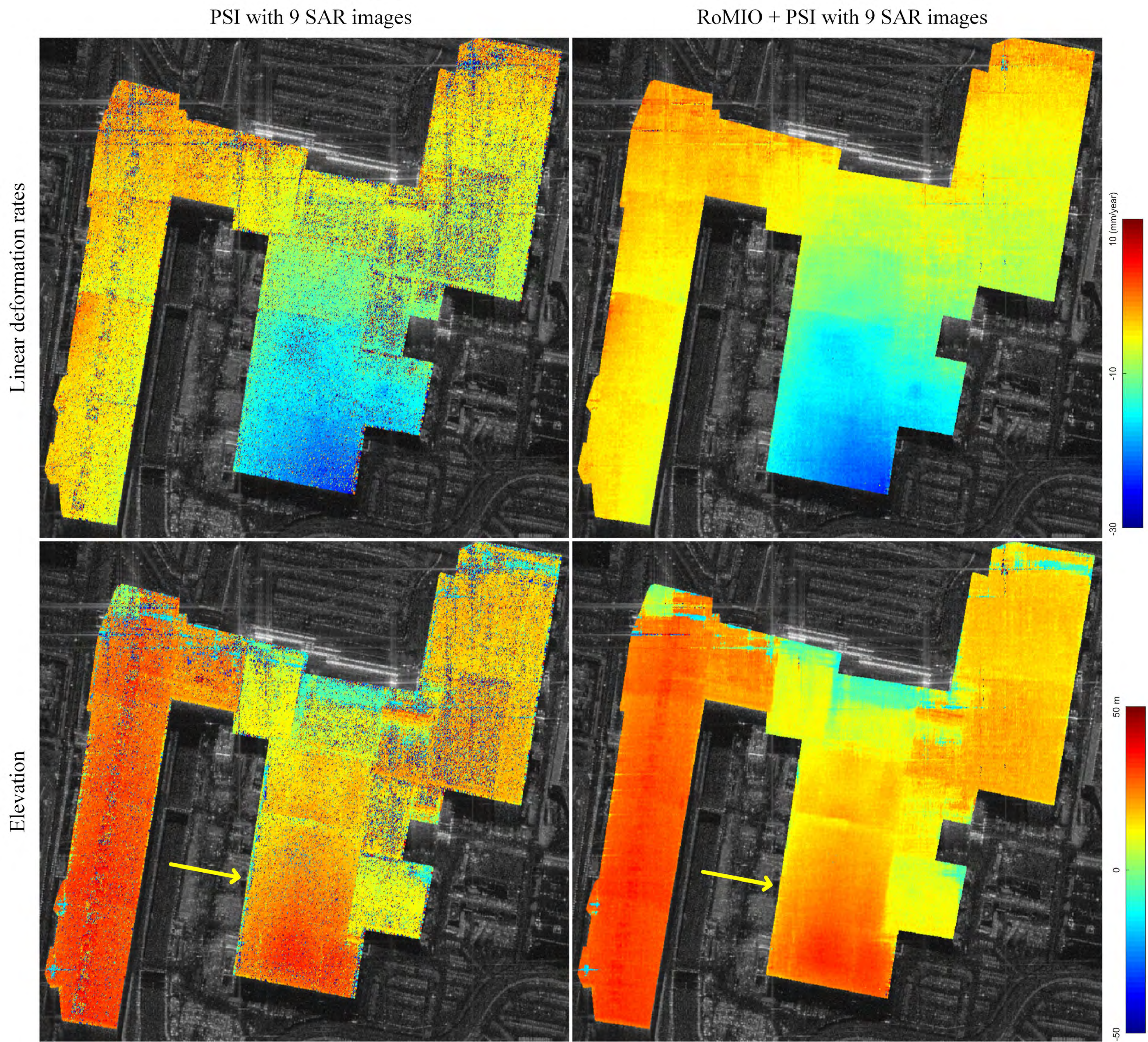}
	\caption{Geophysical parameter estimations (linear deformation rates and elevations) of Las Vegas convention center by PSI and RoMIO + PSI with 9 SAR images (29 images in total). The proposed method can mitigate incorrectly estimated geophysical parameters much better than PSI. Meanwhile, it is worth noting that geometric structures of the building can be preserved well. }
	\label{fg:lasvegas_roof_geo_para_results}
\end{figure*}

\begin{figure*}
	\centering
	\includegraphics[width=\textwidth]{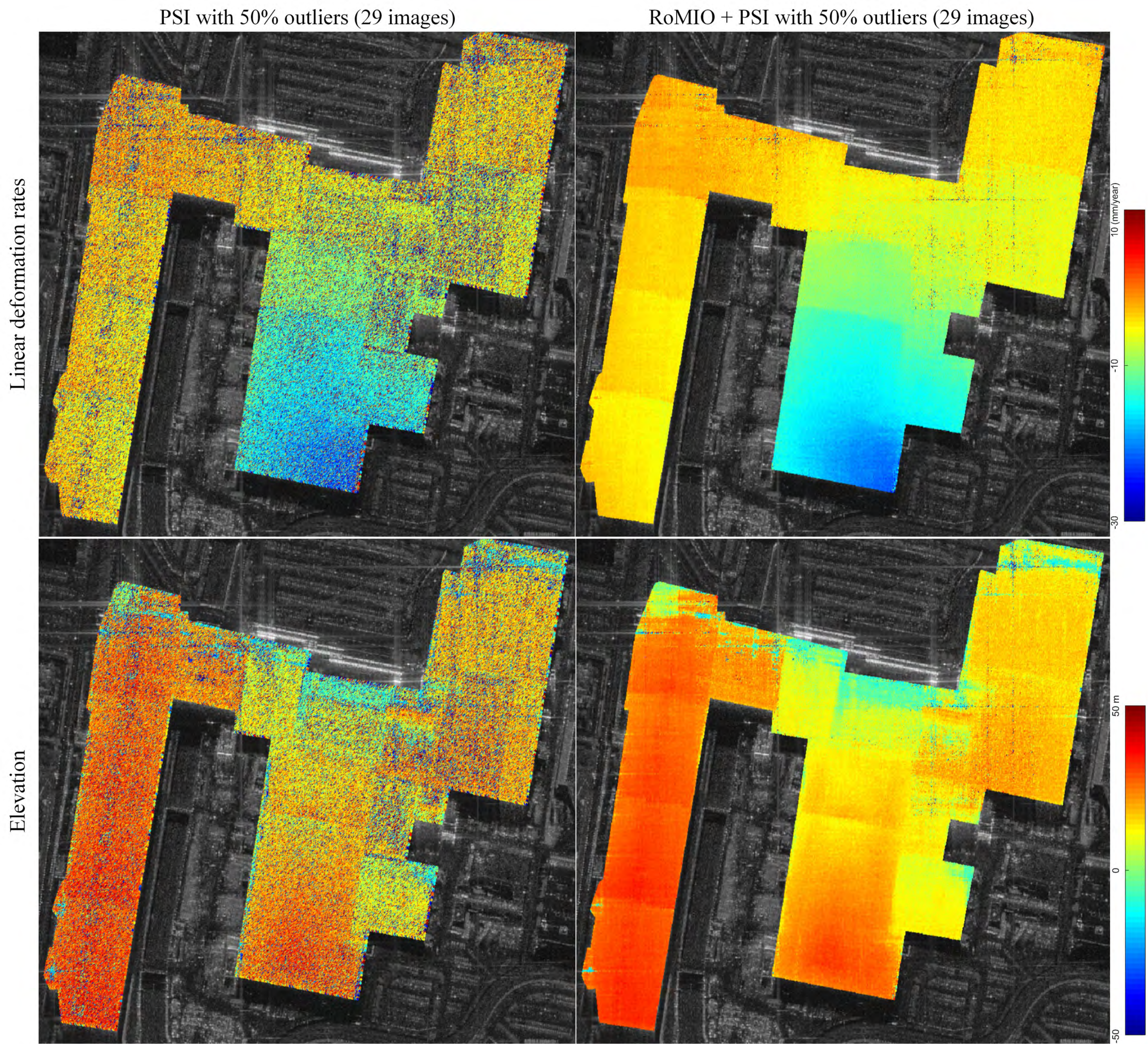}
	\caption{Geophysical parameter estimations (linear deformation rates and elevations) of Las Vegas convention center by PSI and RoMIO + PSI under the stack corrupted by $ 50\% $ outliers. The geometric structures of the building cannot be well interpreted by the results of PSI. In contrast, our method can achieve much more reliable results than PSI.}
	\label{fg:lasvegas_roof_geo_para_adding_outliers}
\end{figure*}

Another TerraSAR-X test dataset is the Las Vegas convention center, as shown in Figure \ref{fg:lasvegas_roof_SAR_google}. The total number of SAR images is $ 29 $. Since the building structure is complex and its spatial area is relatively large ($ 800\times850 $ pixels), we separately processed the four parts of the whole InSAR phase stack as cropped with the red dashed rectangles shown in Figure \ref{fg:lasvegas_roof_SAR_google} (Left). To its right, we also provide the associated optical image from Google Earth. Similar to the previous experiment, we estimate the geophysical parameters by PSI and by the proposed method with a substack ($ 9 $ SAR images), which were selected according to the same baseline criteria described in the previous paragraph. In Figure \ref{fg:lasvegas_2D_baselines}, the 2D distribution of spatial and temporal baselines of the total $ 29 $ measurements is demonstrated, along with those of the selected $ 9 $ measurements for reconstruction. The results are shown in Figure \ref{fg:lasvegas_roof_geo_para_results}. Besides, we manually added $ 50\% $ outliers to the stack and demonstrate the parameters retrieved by PSI and RoMIO + PSI with $ 29 $ SAR images in Figure \ref{fg:lasvegas_roof_geo_para_adding_outliers}.

\begin{figure}
	\centering
	\includegraphics[width=0.5\textwidth]{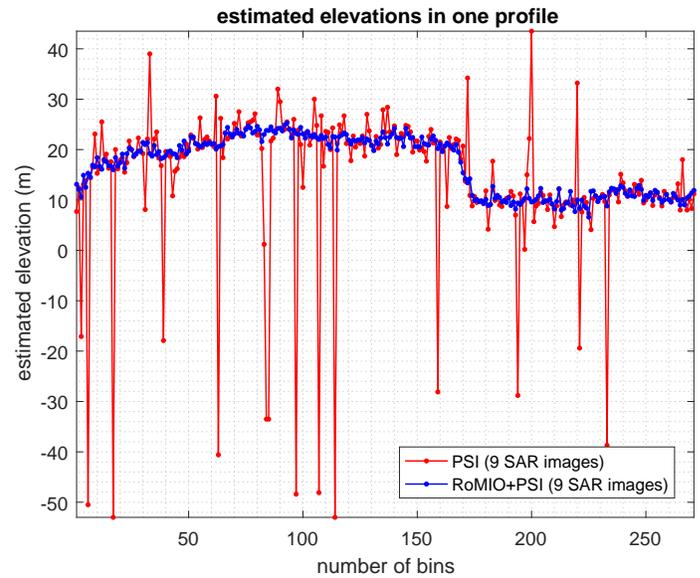}
	\caption{The estimated elevation profiles of the two methods, which are selected by the yellow arrows in Figure \ref{fg:lasvegas_roof_geo_para_results}. The proposed method can preserve resolution by demonstrating a more obvious elevation step jumping than PSI, and simultaneously mitigate incorrectly estimated points.}
	\label{fg:lasvegas_ele_oneprofile}
\end{figure}

\section{Discussions}\label{sc:discussions}
\subsection{Performance in simulations}
According to the results shown in Figure \ref{fg:simu_interferograms_NL_HoRPCA_RIRTD}, although the NL-InSAR result can maintain the smooth fringes very well, the edges of the rectangle in the middle are more blurred compared to the other two results. This can be clearly observed by the plots in Figure \ref{fg:one_profile_simu_inter_phase}. Compared to HoRPCA, the proposed method can better keep the original structure of the interferogram, since it can better capture the low rank structure of the data and model the sparse outliers by enhancing the low rank and the sparsity. Consistently, the evaluation in Table \ref{tb:MSE_performance} shows that under $ 30\% $ percentage of outliers, both NL-InSAR and RoMIO can achieve reliable results. However, when the data is severely corrupted by outliers, e.g. $ 50\% $ outliers, RoMIO can achieve a more robust performance than NL-InSAR.

Combining multipass InSAR techniques, e.g. PSI, with RoMIO can greatly improve the accuracy of parameter estimates. As illustrated in Figure \ref{fg:simu_mountain_realmountain_DeforEle_results}, the results of PSI contain outliers. This is especially true for the result from a subset of the stack. The reason is that the periodogram used PSI is only asymptotically optimal, which means large bias is very likely to occur at low number of images. In contrast, the proposed method can robustly recover the parameters both using the full stack and using a subset of the stack. That is to say the proposed method can in turn effectively reduce the number of images required for a reliable estimation. For the quantitative performance, as illustrated in Table \ref{tb:SD_study_Defor_Ele}, we can see the proposed geophysical parameter retrieval method --- RoMIO + PSI --- can improve the accuracy by a factor of ten to thirty comparing to PSI. This is also transferable to real data, as the simulation closely resembles real TerraSAR-X data. However, some artifacts are observed in the middle of the deformation estimates, which may be caused by choosing a large patch size ($ 128\times128 $) for optimization. Since the spatial information of phase tensors is utilized in the proposed approach, we found that with large patch sizes, over-smoothing artifacts may exist, especially in geometrically complex areas.

As shown in Figure \ref{fg:SD_defor_dif_num_SAR_img}, according to the results of the deformation reconstruction with decreasing numbers of SAR images, the proposed method can achieve a SD around $ 0.3[\mathrm{mm/year}] $, which can improve the estimation accuracy of PSI more than ten times.

Figure \ref{fg:SD_defor_dif_num_SAR_img} shows the accuracy of RoMIO + PSI can maintain at a better and more constant level compared to the PSI whose efficiency decreases linearly w.r.t. the number of images. At the number of  images down to $ 7 $, the accuracy of RoMIO + PSI still keeps at a sub millimeter range which is about 30 times better than PSI. This creates an opportunity of multipass InSAR geophysical parameter reconstruction using very small stacks.

\subsection{Parameter selection}
The two parameters of RoMIO are $ \mu $ and $ \gamma $, where $ \mu $ comes with the introduced Lagrange multiplier term, and $ \gamma $ controls the balance between the low rank tensor $ \mathcal{X} $ and the outlier tensor $ \mathcal{E} $. As introduced in \cite{goldfarb2014robust}, we can keep $ \mu $ constant with the value $ 10\times\mathrm{std}(\mathrm{vec}(\mathcal{G})) $. For tuning $ \gamma $, we first rewrite $ \gamma $ as $ \gamma=\alpha\times\lambda_* $, where a good choice for $ \lambda_* $ can be set as $ \frac{1}{\sqrt{\max(I_1,I_2,\cdots,I_N)}} $ according to \cite{candes2011robust,goldfarb2014robust}, and $ \alpha $ is a factor for tuning. To show the influence of $ \alpha $, Figure \ref{fg:RIRTD_para_setting} presents MSE values of the real-valued residual phases of the phase stack recovered by RoMIO with respect to $ \alpha $ (from $ 0.5\times10^{-3} $ to $ 1\times10^{-2} $), under different percentages of outliers. As shown in the plot, even under a high percentage of outliers, e.g. $ 30\% $, the operable range of $ \alpha $ still keeps relatively wide. Of course, this range decreases as the percentage of outliers increases. Still, the parameter can be tuned using the L-curve method \cite{hansen1993use,kang2017robust}. For a particular dataset, the optimal $ \alpha $ for different percentages of outliers is similar (around $ 5\times10^{-3} $ in our simulation), which means that no assumptions about the amount of outliers is required.

\subsection{Performance in real data}

As shown in Figure \ref{fg:geo_para_berlin_brdige_results}, consistent with the simulations, the proposed method can achieve a more robust estimation result than the classical PSI. In particular, in case of limited number of images, the interpretation of the parameters retrieved by PSI is severely influenced by outliers. The results of the proposed method are more interpretable. One can observe that the amplitudes of the motion tend to increase from one side to the other. One plausible reason is that the deformation allowances on the two sides of the bridge are different. To verify this, a very high resolution image of the bridge is shown in Figure \ref{fg:berlin_bridge_SAR_optical_streetview} (Middle). The yellow ellipses in the image show that there exists certain mechanical clearance between the bridge body and the road it attaches to. Interesting to note is that in the elevation maps in Figure \ref{fg:geo_para_berlin_brdige_results}, there are four elevated regions which correspond to the four lampposts on the bridge. We plot the corresponding profiles of height estimates from the results of PSI and RoMIO + PSI in Figure \ref{fg:berlin_bridge_ele_profiles}. Obviously, the four lampposts are well distinguishable in the result of the proposed method. In order to quantitatively evaluate the performance of the proposed method, the result in Figure \ref{fg:berlin_bridge_ele_profiles} is compared to a centimeter-accuracy LiDAR point cloud shown in Figure \ref{fg:lidar_berlin_bridge}. As InSAR is relative measurement, we robustly adjust the height of bridge surface to match that in the LiDAR point cloud, and then compare the height of lampposts to those in the LiDAR point cloud. To obtain the height of bridge surface as well as the height of lampposts in the two InSAR point clouds and the LiDAR point cloud, we \textit{robustly average} the points within the yellow polygons shown in Figure \ref{fg:geo_para_berlin_brdige_results}, respectively. According to the incidence angle ($ \theta=36.1^{\circ} $), the estimated heights of the four lampposts based on the two methods are shown in Table \ref{fg:LiDAR_Berlin_bridge_lamppost}. On one hand, for such high SNR areas, PSI can achieve a reliable estimation result, while the proposed method indeed increases the height estimates with smaller bias and SD. On the other hand, as illustrated in Figure \ref{fg:berlin_bridge_ele_profiles} and \ref{fg:geo_para_berlin_brdige_results}, for those areas with low SNR such as bridge surface, the proposed method can also obtain much more robust estimates than PSI.

In the results of Las Vegas convention center shown in Figure \ref{fg:lasvegas_roof_geo_para_results}, the proposed method can mitigate the incorrectly estimated geophysical parameters much better than PSI, under limited SAR images. Besides, it is worth noting that the geometric structure of the object can be well preserved. For instance, as shown in Figure \ref{fg:lasvegas_ele_oneprofile}, we plot the elevation profiles indicated by the yellow arrows from the two results in Figure \ref{fg:lasvegas_roof_geo_para_results}. The proposed method can preserve resolution by displaying a more obvious elevation step jumping than PSI, and simultaneously mitigates outliers. Moreover, when we synthetically corrupt the data by $ 50\% $ of outliers, the geometric structures of the building cannot be well interpreted by the results of PSI as shown in Figure \ref{fg:lasvegas_roof_geo_para_adding_outliers}. In contrast, the proposed method can achieve much more reliable results.

\section{Conclusion}\label{sc:conclustion}
This paper studied the low rank property of object-based InSAR phase stacks and proposed Robust Multipass InSAR technique via Object-based low rank tensor decomposition (RoMIO). RoMIO can be combined with conventional multipass InSAR techniques to improve the estimation accuracy of geophysical parameters. Taking PSI as an example, the paper demonstrated that in typical condition of very high resolution spaceborne InSAR data, e.g. object size of $ 10 \mathrm{m} $, $ 5 $dB SNR and $ 10-20 $ SAR images, the proposed approach can improve the estimation accuracy of geophysical parameters by a factor of ten to thirty, especially in the presence of outliers. These merits can in turn efficiently reduce the number of SAR images for a reliable estimation.

Based on our experiments, we can see that the spatial sizes of tensors can influence the efficiency of the proposed method. On one hand, with large spatial sizes, the low rank property of the phase tensor is generally more prominent, which gives a wide operable range of the regularization parameters in the optimization. But, over-smoothing artifacts may exist, especially in some geometrically complex areas. On the other hand, with small spatial sizes, although it can be a benefit for preserving small detail, the regularization parameters must be carefully tuned. Otherwise, the reconstructed phase tensor may have the risk to be turned into a rank-1 tensor. Therefore, the tensor size should be large enough to promote low-rankness of the true phase and the sparsity of outliers, but small enough to exclude complicated structures. According to the experiments in the paper, the typical patch size we utilized is around $ 100\times100 $ pixels, and this can be improved by exploiting adaptive window.

Besides, the proposed approach is suitable for operational processing, as the only parameter that needs to be tuned, i.e. $ \alpha $, was shown to usually lie in the range from $ 1\times10^{-3} $ to $ 1\times10^{-1} $ based on both the simulated and real data experiments. Besides, the approach can easily be parallelized by carrying it out patch-wisely.

Currently, this approach relies on the segmentation of objects. For future work, we would like to investigate lower level geometric information in SAR images to relax the requirement of object masks. Furthermore, we are also planning to research objects in non-urban areas, where few of them present regular shapes, and attempt to investigate their inherent property which can be utilized for the improvement of geophysical parameter retrieval, based on the geometric information.

\section*{Acknowledgment}
The authors would like to thank H. Hirschm\"uller of DLR Institute of Robotics and Mechatronics for providing the orthorectified optical image of Berlin.

The authors would like to thank the reviewers for their valuable suggestions.

\bibliographystyle{IEEEtran}
\bibliography{bibliography}

 \newcommand{\noop}[1]{}
\begin{thebibliography}{10}
\providecommand{\url}[1]{#1}
\csname url@samestyle\endcsname
\providecommand{\newblock}{\relax}
\providecommand{\bibinfo}[2]{#2}
\providecommand{\BIBentrySTDinterwordspacing}{\spaceskip=0pt\relax}
\providecommand{\BIBentryALTinterwordstretchfactor}{4}
\providecommand{\BIBentryALTinterwordspacing}{\spaceskip=\fontdimen2\font plus
\BIBentryALTinterwordstretchfactor\fontdimen3\font minus
  \fontdimen4\font\relax}
\providecommand{\BIBforeignlanguage}[2]{{%
\expandafter\ifx\csname l@#1\endcsname\relax
\typeout{** WARNING: IEEEtran.bst: No hyphenation pattern has been}%
\typeout{** loaded for the language `#1'. Using the pattern for}%
\typeout{** the default language instead.}%
\else
\language=\csname l@#1\endcsname
\fi
#2}}
\providecommand{\BIBdecl}{\relax}
\BIBdecl

\bibitem{kang2017robust}
J.~Kang, Y.~Wang, M.~K{\"o}rner, and X.~X. Zhu, ``{Robust Object-Based
  Multipass InSAR Deformation Reconstruction},'' \emph{IEEE Transactions on
  Geoscience and Remote Sensing}, no.~99, pp. 1--13, 2017.

\bibitem{ferretti2001permanent}
A.~Ferretti, C.~Prati, and F.~Rocca, ``{Permanent scatterers in SAR
  interferometry},'' \emph{IEEE Transactions on geoscience and remote sensing},
  vol.~39, no.~1, pp. 8--20, 2001.

\bibitem{adam2003development}
N.~Adam, B.~Kampes, M.~Eineder, J.~Worawattanamateekul, and M.~Kircher, ``The
  development of a scientific permanent scatterer system,'' in \emph{ISPRS
  Workshop High Resolution Mapping from Space, Hannover, Germany}, vol. 2003,
  2003, p.~6.

\bibitem{fornaro2009deformation}
G.~Fornaro, A.~Pauciullo, and F.~Serafino, ``Deformation monitoring over large
  areas with multipass differential sar interferometry: a new approach based on
  the use of spatial differences,'' \emph{International Journal of Remote
  Sensing}, vol.~30, no.~6, pp. 1455--1478, 2009.

\bibitem{sousa2011persistent}
J.~J. Sousa, A.~J. Hooper, R.~F. Hanssen, L.~C. Bastos, and A.~M. Ruiz,
  ``{Persistent scatterer InSAR: a comparison of methodologies based on a model
  of temporal deformation vs. spatial correlation selection criteria},''
  \emph{Remote Sensing of Environment}, vol. 115, no.~10, pp. 2652--2663, 2011.

\bibitem{gernhardt2012deformation}
S.~Gernhardt and R.~Bamler, ``{Deformation monitoring of single buildings using
  meter-resolution SAR data in PSI},'' \emph{ISPRS journal of photogrammetry
  and remote sensing}, vol.~73, pp. 68--79, 2012.

\bibitem{kampes2006radar}
B.~M. Kampes, \emph{Radar interferometry}.\hskip 1em plus 0.5em minus
  0.4em\relax Dordrecht, The Netherlands: Springer, 2006.

\bibitem{wang2014efficient}
Y.~Wang, X.~X. Zhu, and R.~Bamler, ``An efficient tomographic inversion
  approach for urban mapping using meter resolution {SAR} image stacks,''
  \emph{IEEE Geoscience and Remote Sensing Letters}, vol.~11, no.~7, pp.
  1250--1254, 2014.

\bibitem{costantini2014persistent}
M.~Costantini, S.~Falco, F.~Malvarosa, F.~Minati, F.~Trillo, and F.~Vecchioli,
  ``{Persistent scatterer pair interferometry: approach and application to
  COSMO-SkyMed SAR data},'' \emph{IEEE Journal of Selected Topics in Applied
  Earth Observations and Remote Sensing}, vol.~7, no.~7, pp. 2869--2879, 2014.

\bibitem{zhang2011modeling}
L.~Zhang, X.~Ding, and Z.~Lu, ``{Modeling PSInSAR time series without phase
  unwrapping},'' \emph{IEEE Transactions on Geoscience and Remote Sensing},
  vol.~49, no.~1, pp. 547--556, 2011.

\bibitem{de2009detection}
A.~De~Maio, G.~Fornaro, and A.~Pauciullo, ``{Detection of single scatterers in
  multidimensional SAR imaging},'' \emph{IEEE Transactions on Geoscience and
  Remote Sensing}, vol.~47, no.~7, pp. 2284--2297, 2009.

\bibitem{ferretti2011new}
A.~Ferretti, A.~Fumagalli, F.~Novali, C.~Prati, F.~Rocca, and A.~Rucci, ``{A
  new algorithm for processing interferometric data-stacks: SqueeSAR},''
  \emph{IEEE Transactions on Geoscience and Remote Sensing}, vol.~49, no.~9,
  pp. 3460--3470, 2011.

\bibitem{goel2012advanced}
K.~Goel and N.~Adam, ``An advanced algorithm for deformation estimation in
  non-urban areas,'' \emph{ISPRS journal of photogrammetry and remote sensing},
  vol.~73, pp. 100--110, 2012.

\bibitem{Wang201289}
Y.~Wang, X.~X. Zhu, and R.~Bamler, ``Retrieval of phase history parameters from
  distributed scatterers in urban areas using very high resolution {SAR}
  data,'' \emph{{ISPRS} Journal of Photogrammetry and Remote Sensing}, vol.~73,
  pp. 89 -- 99, 2012.

\bibitem{jiang2015fast}
M.~Jiang, X.~Ding, R.~F. Hanssen, R.~Malhotra, and L.~Chang, ``Fast
  statistically homogeneous pixel selection for covariance matrix estimation
  for multitemporal insar,'' \emph{IEEE Transactions on Geoscience and Remote
  Sensing}, vol.~53, no.~3, pp. 1213--1224, 2015.

\bibitem{samiei2016phase}
S.~Samiei-Esfahany, J.~E. Martins, F.~van Leijen, and R.~F. Hanssen, ``{Phase
  Estimation for Distributed Scatterers in InSAR Stacks Using Integer Least
  Squares Estimation},'' \emph{IEEE Transactions on Geoscience and Remote
  Sensing}, vol.~54, no.~10, pp. 5671--5687, 2016.

\bibitem{wang2016robust}
Y.~Wang and X.~X. Zhu, ``{Robust estimators for multipass SAR
  interferometry},'' \emph{IEEE Transactions on Geoscience and Remote Sensing},
  vol.~54, no.~2, pp. 968--980, 2016.

\bibitem{Fornaro2003}
G.~Fornaro, F.~Serafino, and F.~Soldovieri, ``{Three-dimensional focusing with
  multipass SAR data},'' \emph{IEEE Transactions on Geoscience and Remote
  Sensing}, vol.~41, no.~3, pp. 507--517, 2003.

\bibitem{lombardini2005}
F.~Lombardini, ``{Differential tomography: a new framework for SAR
  interferometry},'' \emph{IEEE Transactions on Geoscience and Remote Sensing},
  vol.~43, no.~1, pp. 37--44, Jan 2005.

\bibitem{zhu2010very}
X.~X. Zhu and R.~Bamler, ``{Very high resolution spaceborne SAR tomography in
  urban environment},'' \emph{IEEE Transactions on Geoscience and Remote
  Sensing}, vol.~48, no.~12, pp. 4296--4308, 2010.

\bibitem{zhunolinear2011}
------, ``{Let's Do the Time Warp: Multicomponent Nonlinear Motion Estimation
  in Differential SAR Tomography},'' \emph{IEEE Geoscience and Remote Sensing
  Letters}, vol.~8, no.~4, pp. 735--739, July 2011.

\bibitem{reale2011}
D.~Reale, G.~Fornaro, A.~Pauciullo, X.~Zhu, and R.~Bamler, ``{Tomographic
  Imaging and Monitoring of Buildings With Very High Resolution SAR Data},''
  \emph{IEEE Geoscience and Remote Sensing Letters}, vol.~8, no.~4, pp.
  661--665, July 2011.

\bibitem{fornaro2014tomographic}
G.~Fornaro, F.~Lombardini, A.~Pauciullo, D.~Reale, and F.~Viviani,
  ``{Tomographic processing of interferometric SAR data: Developments,
  applications, and future research perspectives},'' \emph{IEEE Signal
  Processing Magazine}, vol.~31, no.~4, pp. 41--50, 2014.

\bibitem{zhu2010tomographic}
X.~X. Zhu and R.~Bamler, ``{Tomographic SAR inversion by L1 norm
  regularization---The compressive sensing approach},'' \emph{IEEE Transactions
  on Geoscience and Remote Sensing}, vol.~48, no.~10, pp. 3839--3846, 2010.

\bibitem{budillon2011three}
A.~Budillon, A.~Evangelista, and G.~Schirinzi, ``{Three-dimensional SAR
  focusing from multipass signals using compressive sampling},'' \emph{IEEE
  Transactions on Geoscience and Remote Sensing}, vol.~49, no.~1, pp. 488--499,
  2011.

\bibitem{zhu2014superresolving}
X.~X. Zhu and R.~Bamler, ``{Superresolving SAR tomography for multidimensional
  imaging of urban areas: compressive sensing-based tomoSAR inversion},''
  \emph{IEEE Signal Processing Magazine}, vol.~31, no.~4, pp. 51--58, 2014.

\bibitem{eineder2011imaging}
M.~Eineder, C.~Minet, P.~Steigenberger, X.~Cong, and T.~Fritz, ``{Imaging
  geodesy---Toward centimeter-level ranging accuracy with TerraSAR-X},''
  \emph{IEEE Transactions on Geoscience and Remote Sensing}, vol.~49, no.~2,
  pp. 661--671, 2011.

\bibitem{gisinger2015precise}
C.~Gisinger, U.~Balss, R.~Pail, X.~X. Zhu, S.~Montazeri, S.~Gernhardt, and
  M.~Eineder, ``{Precise three-dimensional stereo localization of corner
  reflectors and persistent scatterers with TerraSAR-X},'' \emph{IEEE
  Transactions on Geoscience and Remote Sensing}, vol.~53, no.~4, pp.
  1782--1802, 2015.

\bibitem{zhu2016geodetic}
X.~X. Zhu, S.~Montazeri, C.~Gisinger, R.~F. Hanssen, and R.~Bamler, ``{Geodetic
  SAR tomography},'' \emph{IEEE Transactions on Geoscience and Remote Sensing},
  vol.~54, no.~1, pp. 18--35, 2016.

\bibitem{cao2016phase}
N.~Cao, H.~Lee, and H.~C. Jung, ``{A phase-decomposition-based PSInSAR
  processing method},'' \emph{IEEE Transactions on Geoscience and Remote
  Sensing}, vol.~54, no.~2, pp. 1074--1090, 2016.

\bibitem{schmitt2014adaptive}
M.~Schmitt, J.~L. Sch{\"o}nberger, and U.~Stilla, ``{Adaptive covariance matrix
  estimation for multi-baseline InSAR data stacks},'' \emph{IEEE Transactions
  on Geoscience and Remote Sensing}, vol.~52, no.~11, pp. 6807--6817, 2014.

\bibitem{schmitt2014adaptivemultilooking}
M.~Schmitt and U.~Stilla, ``{Adaptive multilooking of airborne single-pass
  multi-baseline InSAR stacks},'' \emph{IEEE Transactions on Geoscience and
  Remote Sensing}, vol.~52, no.~1, pp. 305--312, 2014.

\bibitem{fornaro2015caesar}
G.~Fornaro, S.~Verde, D.~Reale, and A.~Pauciullo, ``{CAESAR: An approach based
  on covariance matrix decomposition to improve multibaseline--multitemporal
  interferometric SAR processing},'' \emph{IEEE Transactions on Geoscience and
  Remote Sensing}, vol.~53, no.~4, pp. 2050--2065, 2015.

\bibitem{neumann2010estimation}
M.~Neumann, L.~Ferro-Famil, and A.~Reigber, ``{Estimation of forest structure,
  ground, and canopy layer characteristics from multibaseline polarimetric
  interferometric SAR data},'' \emph{IEEE Transactions on Geoscience and Remote
  Sensing}, vol.~48, no.~3, pp. 1086--1104, 2010.

\bibitem{tebaldini2010single}
S.~Tebaldini, ``{Single and multipolarimetric SAR tomography of forested areas:
  A parametric approach},'' \emph{IEEE Transactions on Geoscience and Remote
  Sensing}, vol.~48, no.~5, pp. 2375--2387, 2010.

\bibitem{schmitt2014maximum}
M.~Schmitt and U.~Stilla, ``Maximum-likelihood-based approach for single-pass
  synthetic aperture radar tomography over urban areas,'' \emph{IET Radar,
  Sonar \& Navigation}, vol.~8, no.~9, pp. 1145--1153, 2014.

\bibitem{deledalle2011nl}
C.-A. Deledalle, L.~Denis, and F.~Tupin, ``{Nl-insar: Nonlocal interferogram
  estimation},'' \emph{IEEE Transactions on Geoscience and Remote Sensing},
  vol.~49, no.~4, pp. 1441--1452, 2011.

\bibitem{zhu2014improving}
X.~X. Zhu, R.~Bamler, M.~Lachaise, F.~Adam, Y.~Shi, and M.~Eineder,
  ``{Improving TanDEM-X DEMs by non-local InSAR filtering},'' in \emph{EUSAR
  2014; 10th European Conference on Synthetic Aperture Radar; Proceedings
  of}.\hskip 1em plus 0.5em minus 0.4em\relax VDE, 2014, pp. 1--4.

\bibitem{deledalle2014exploiting}
C.-A. Deledalle, L.~Denis, G.~Poggi, F.~Tupin, and L.~Verdoliva, ``Exploiting
  patch similarity for sar image processing: the nonlocal paradigm,''
  \emph{IEEE Signal Processing Magazine}, vol.~31, no.~4, pp. 69--78, 2014.

\bibitem{sica2015nonlocal}
F.~Sica, D.~Reale, G.~Poggi, L.~Verdoliva, and G.~Fornaro, ``{Nonlocal Adaptive
  Multilooking in SAR Multipass Differential Interferometry},'' \emph{IEEE
  Journal of Selected Topics in Applied Earth Observations and Remote Sensing},
  vol.~8, no.~4, pp. 1727--1742, 2015.

\bibitem{zhumslimmer}
X.~X. Zhu, N.~Ge, and M.~Shahzad, ``{Joint Sparsity in SAR Tomography for Urban
  Mapping},'' \emph{IEEE Journal of Selected Topics in Signal Processing},
  vol.~9, no.~8, pp. 1498--1509, Dec 2015.

\bibitem{kang2016object}
J.~Kang, Y.~Wang, M.~K{\"o}rner, and X.~X. Zhu, ``{Object-based InSAR
  deformation reconstruction with application to bridge monitoring},'' in
  \emph{Geoscience and Remote Sensing Symposium (IGARSS), 2016 IEEE
  International}.\hskip 1em plus 0.5em minus 0.4em\relax IEEE, 2016, pp.
  6871--6874.

\bibitem{zhou2015low}
X.~Zhou, C.~Yang, H.~Zhao, and W.~Yu, ``Low-rank modeling and its applications
  in image analysis,'' \emph{ACM Computing Surveys (CSUR)}, vol.~47, no.~2,
  p.~36, 2015.

\bibitem{jolliffe2002principal}
I.~Jolliffe, \emph{Principal component analysis}.\hskip 1em plus 0.5em minus
  0.4em\relax Wiley Online Library, 2002.

\bibitem{yousif2013improving}
O.~Yousif and Y.~Ban, ``Improving urban change detection from multitemporal sar
  images using pca-nlm,'' \emph{IEEE Transactions on Geoscience and Remote
  Sensing}, vol.~51, no.~4, pp. 2032--2041, 2013.

\bibitem{chen2011denoising}
G.~Chen and S.-E. Qian, ``Denoising of hyperspectral imagery using principal
  component analysis and wavelet shrinkage,'' \emph{IEEE Transactions on
  Geoscience and remote sensing}, vol.~49, no.~3, pp. 973--980, 2011.

\bibitem{yao2003genetic}
H.~Yao and L.~Tian, ``A genetic-algorithm-based selective principal component
  analysis (ga-spca) method for high-dimensional data feature extraction,''
  \emph{IEEE Transactions on Geoscience and Remote Sensing}, vol.~41, no.~6,
  pp. 1469--1478, 2003.

\bibitem{candes2011robust}
E.~J. Cand{\`e}s, X.~Li, Y.~Ma, and J.~Wright, ``Robust principal component
  analysis?'' \emph{Journal of the ACM (JACM)}, vol.~58, no.~3, p.~11, 2011.

\bibitem{zhang2014hyperspectral}
H.~Zhang, W.~He, L.~Zhang, H.~Shen, and Q.~Yuan, ``Hyperspectral image
  restoration using low-rank matrix recovery,'' \emph{IEEE Transactions on
  Geoscience and Remote Sensing}, vol.~52, no.~8, pp. 4729--4743, 2014.

\bibitem{borcea2013synthetic}
L.~Borcea, T.~Callaghan, and G.~Papanicolaou, ``Synthetic aperture radar
  imaging and motion estimation via robust principal component analysis,''
  \emph{SIAM Journal on Imaging Sciences}, vol.~6, no.~3, pp. 1445--1476, 2013.

\bibitem{goldfarb2014robust}
D.~Goldfarb and Z.~Qin, ``Robust low-rank tensor recovery: Models and
  algorithms,'' \emph{SIAM Journal on Matrix Analysis and Applications},
  vol.~35, no.~1, pp. 225--253, 2014.

\bibitem{kolda2009tensor}
T.~G. Kolda and B.~W. Bader, ``Tensor decompositions and applications,''
  \emph{SIAM review}, vol.~51, no.~3, pp. 455--500, 2009.

\bibitem{cichocki2015tensor}
A.~Cichocki, D.~Mandic, L.~De~Lathauwer, G.~Zhou, Q.~Zhao, C.~Caiafa, and H.~A.
  Phan, ``Tensor decompositions for signal processing applications: From
  two-way to multiway component analysis,'' \emph{IEEE Signal Processing
  Magazine}, vol.~32, no.~2, pp. 145--163, 2015.

\bibitem{de1994singular}
L.~De~Lathauwer, B.~De~Moor, J.~Vandewalle, and B.~S.~S. by~Higher-Order,
  ``Singular value decomposition,'' in \emph{Proc. EUSIPCO-94, Edinburgh,
  Scotland, UK}, vol.~1, 1994, pp. 175--178.

\bibitem{de2000multilinear}
L.~De~Lathauwer, B.~De~Moor, and J.~Vandewalle, ``A multilinear singular value
  decomposition,'' \emph{SIAM journal on Matrix Analysis and Applications},
  vol.~21, no.~4, pp. 1253--1278, 2000.

\bibitem{candes2008enhancing}
E.~J. Candes, M.~B. Wakin, and S.~P. Boyd, ``{Enhancing sparsity by reweighted
  L1 minimization},'' \emph{Journal of Fourier analysis and applications},
  vol.~14, no. 5-6, pp. 877--905, 2008.

\bibitem{peng2014reweighted}
Y.~Peng, J.~Suo, Q.~Dai, and W.~Xu, ``Reweighted low-rank matrix recovery and
  its application in image restoration,'' \emph{IEEE transactions on
  cybernetics}, vol.~44, no.~12, pp. 2418--2430, 2014.

\bibitem{boyd2011distributed}
S.~Boyd, N.~Parikh, E.~Chu, B.~Peleato, and J.~Eckstein, ``Distributed
  optimization and statistical learning via the alternating direction method of
  multipliers,'' \emph{Foundations and Trends{\textregistered} in Machine
  Learning}, vol.~3, no.~1, pp. 1--122, 2011.

\bibitem{gu2014weighted}
S.~Gu, L.~Zhang, W.~Zuo, and X.~Feng, ``Weighted nuclear norm minimization with
  application to image denoising,'' in \emph{Proceedings of the IEEE Conference
  on Computer Vision and Pattern Recognition}, 2014, pp. 2862--2869.

\bibitem{hansen1993use}
P.~C. Hansen and D.~P. O'Leary, ``{The use of the L-curve in the regularization
  of discrete ill-posed problems},'' \emph{SIAM Journal on Scientific
  Computing}, vol.~14, no.~6, pp. 1487--1503, 1993.

\bibitem{hirschmuller2008stereo}
H.~Hirschmuller, ``Stereo processing by semiglobal matching and mutual
  information,'' \emph{IEEE Transactions on pattern analysis and machine
  intelligence}, vol.~30, no.~2, pp. 328--341, 2008.

\end{thebibliography}

\begin{IEEEbiography}[{\includegraphics[width=1in,height=1.25in,clip,keepaspectratio]{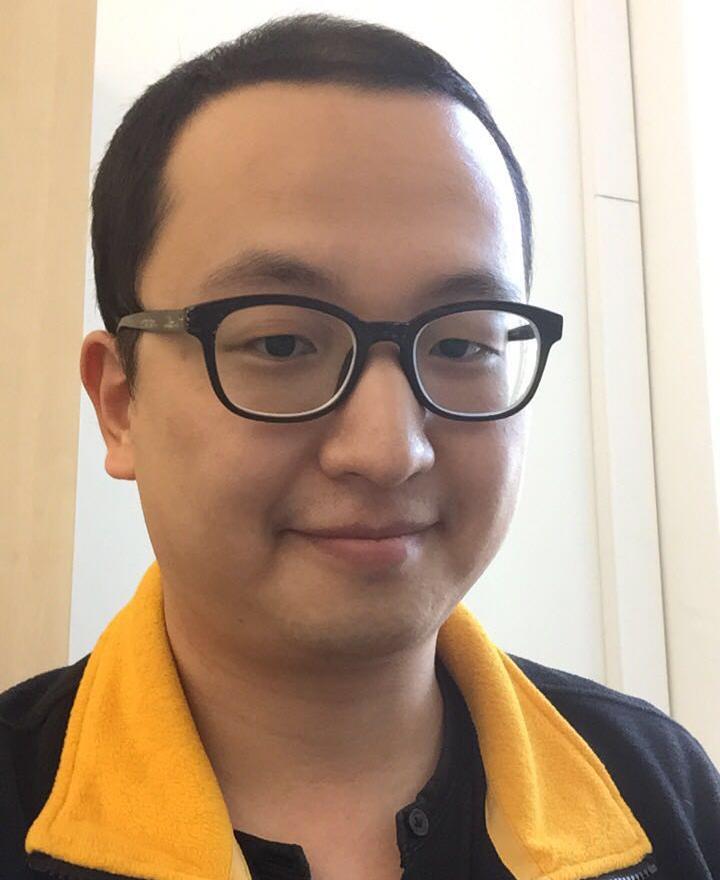}}]{Jian Kang}
	(S'16) received B.S. and M.E. degrees in electronic engineering from Harbin Institute of Technology (HIT), Harbin, China, in 2013 and 2015, respectively. Since 2015, he has been pursuing Doctoral degree with the Chair of Signal Processing in Earth Observation (SiPEO), Technical University of Munich (TUM), Munich, Germany. His research interests include multi-dimensional data analysis, geophysical parameter estimation based on InSAR data and machine learning in optical images.

\end{IEEEbiography}

\begin{IEEEbiography}[{\includegraphics[width=1in,height=1.25in,clip,keepaspectratio]{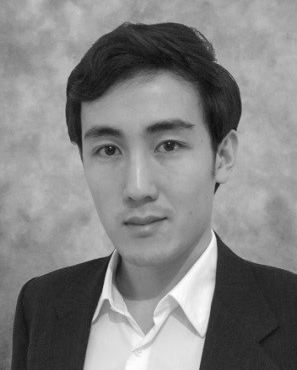}}]{Yuanyuan Wang}
	(S'10-M'14) received his B.Eng. (Hons.) degree in electrical engineering from The Hong Kong Polytechnic University, Hong Kong, China, in 2008, and the M.Sc. degree as well as the Dr.-Ing. degree from Technical University of Munich (TUM), Munich, Germany in 2010 and 2015, respectively. In June and July of 2014, he was a guest scientist at the Institute of Visual Computing, ETH Zurich, Switzerland. Currently, he is with Signal Processing in Earth Observation (SiPEO http://www.sipeo.bgu.tum.de/), TUM.
	His research interests include optimal and robust parameters estimation in multibaseline InSAR techniques, multisensor fusion algorithms of SAR and optical data, non-linear optimization with complex numbers, and the applications of these techniques in urban and volcanic areas. He was one of the best reviewers of IEEE TGRS 2016.
\end{IEEEbiography}

\begin{IEEEbiography}[{\includegraphics[width=1in,height=1.25in,clip,keepaspectratio]{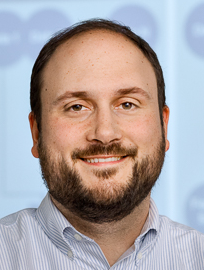}}]{Michael Schmitt}
	(S'08-M'14-SM'16) received the Dipl.-Ing. degree in geodesy and geoinformation and the Dr.-Ing. degree in remote sensing from the Technical University of Munich (TUM), Munich, Germany, in 2009 and 2014, respectively. Since 2015, he has been a Senior Researcher and the Deputy Head at the Professorship for Signal Processing in Earth Observation at TUM. In 2016, he was a Guest Scientist at the University of Massachusetts Amherst, Amherst, MA, USA. His research focuses on signal and image processing for the extraction of information from remote sensing data. In particular, he is interested in sensor data fusion with emphasis on the joint exploitation of optical and radar data; in 3D reconstruction by techniques such as SAR interferometry, SAR tomography, radargrammetry, or photogrammetry; and in millimeter wave SAR remote sensing. Dr. Schmitt is a Co-Chair of the International Society for Photogrammetry and Remote Sensing Working Group I/3 on SAR and Microwave Sensing and frequently serves as a reviewer for a number of renowned international journals. In 2013 and 2015, he was elected the IEEE Geoscience and Remote Sensing Letters Best Reviewer, leading to his appointment as an Associate Editor of the journal in 2016.
\end{IEEEbiography}

\begin{IEEEbiography}[{\includegraphics[width=1in,height=1.25in,clip,keepaspectratio]{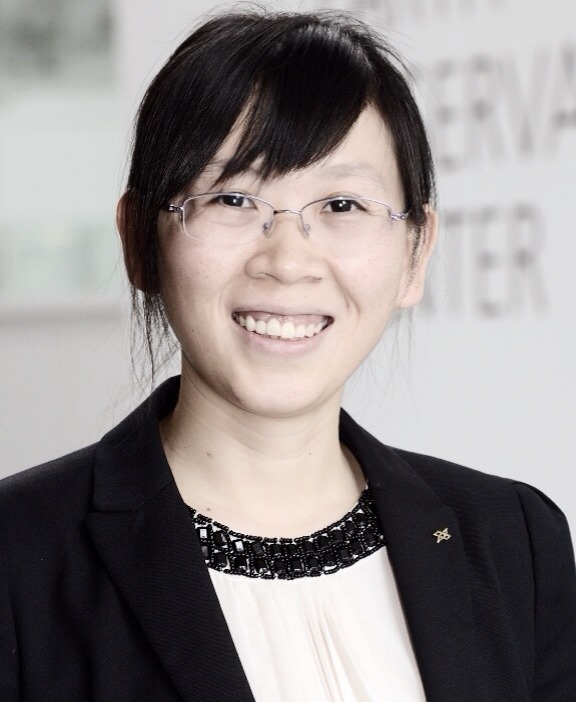}}]{Xiao Xiang Zhu}
	(S'10-M'12-SM'14) received the Master (M.Sc.) degree, her doctor of engineering (Dr.-Ing.) degree and her ``Habilitation'' in the field of signal processing from Technical University of Munich (TUM), Munich, Germany, in 2008, 2011 and 2013, respectively.
\par
She is currently the Professor for Signal Processing in Earth Observation (www.sipeo.bgu.tum.de) at Technical University of Munich (TUM) and German Aerospace Center (DLR); the head of the Team Signal Analysis at DLR; and the head of the Helmholtz Young Investigator Group "SiPEO" at DLR and TUM. Prof. Zhu was a guest scientist or visiting professor at the Italian National Research Council (CNR-IREA), Naples, Italy, Fudan University, Shanghai, China, the University of Tokyo, Tokyo, Japan and University of California, Los Angeles, United States in 2009, 2014, 2015 and 2016, respectively. Her main research interests are remote sensing and Earth observation, signal processing, machine learning and data science, with a special application focus on global urban mapping.
\par
Dr. Zhu is a member of young academy (Junge Akademie/Junges Kolleg) at the Berlin-Brandenburg Academy of Sciences and Humanities and the German National Academy of Sciences Leopoldina and the Bavarian Academy of Sciences and Humanities. She is an associate Editor of IEEE Transactions on Geoscience and Remote Sensing.
\end{IEEEbiography}

\end{document}